\documentclass{iopart}

\usepackage{epsfig}
\usepackage{graphics}
\usepackage{graphicx}
\usepackage{comment}
\usepackage{amsfonts}
\expandafter\let\csname equation*\endcsname\relax
\expandafter\let\csname endequation*\endcsname\relax
\usepackage{amsmath, amssymb}
\usepackage{bm}
\usepackage{xcolor,xspace}
\usepackage{wasysym}
\usepackage{times}
\usepackage{mathptmx}
\usepackage{appendix}
\usepackage{mathrsfs}
\usepackage{lipsum}
\usepackage[colorlinks]{hyperref}
\usepackage[all]{hypcap}
\usepackage{array}
\usepackage{caption}

\usepackage{algorithm}
\usepackage{algpseudocode} 
\usepackage{setspace} 

\def\bx{ {\bm x} }

\def\vecd{ {\vec d} }
\def\vecD{ {\vec D} }
\def\vecx{ {\vec x} }
\def\vecX{ {\vec X} }
\def\vecy{ {\vec y} }
\def\vecY{ {\vec Y} }

\begin{document}

\newcommand{\LIGOLab}{LIGO Laboratory, California Institute of Technology, MS 100-36, Pasadena, California 91125, USA}
\newcommand{\TAPIR}{Theoretical Astrophysics, Walter Burke Institute for Theoretical Physics, California Institute of Technology, Pasadena, California 91125, USA}
\newcommand{\Radboud}{Department of Astrophysics/IMAPP, Radboud University Nijmegen, P.O. Box 9010, 6525 GL Nijmegen, The Netherlands }
\newcommand{\contribution}{Both authors have contributed equally}

\begin{flushright}
LIGO-P1600064
\end{flushright}

\title[Reduced-order spline interpolation]{Fast and efficient evaluation of gravitational waveforms via reduced-order spline interpolation}

\author{Chad R.~Galley\textsuperscript{1 *} and Patricia Schmidt\textsuperscript{1,2,3 *}}
\address{\textsuperscript{1} \TAPIR}
\address{\textsuperscript{2} \LIGOLab}
\address{\textsuperscript{3} \Radboud}
\address{\textsuperscript{*} \contribution}
\eads{\mailto{crgalley@tapir.caltech.edu}, ~\mailto{P.Schmidt@astro.ru.nl}}

\begin{abstract}
Numerical simulations of merging black hole binaries produce the most accurate
gravitational waveforms.
The availability of hundreds of these numerical relativity (NR) waveforms, often containing many higher spherical
harmonic modes, allows one to study many aspects of gravitational waves.
Amongst these are the response of
data analysis pipelines, the calibration of semi-analytical models, the building of reduced-order surrogates,
the estimation of the parameters of detected gravitational waves, and the composition of public catalogs of NR waveform data. 
The large number of generated NR waveforms consequently requires efficient data storage and handling, 
especially since many more waveforms will be generated at an increased rate in the forthcoming years. 
In addition, gravitational wave data analyses often require the NR waveforms to be 
interpolated and uniformly resampled at high sampling rates. 
Previously, this resulted in very large data files (up to $\sim$ several GB) in memory-intensive operations, which is unfeasible when confronted with hundreds of multi-modal NR waveforms. 
To handle these challenges, we present a simple and efficient method to significantly \emph{compress} the original waveform data sets
while accurately reproducing the original data via spline interpolation. The method is generically applicable to relatively smooth, one-dimensional datasets and uses a greedy algorithm 
to determine the most relevant subset of the full data such that a spline interpolant of
a specified polynomial degree will represent the original data to within a requested
point-wise tolerance. 
We find significant compression of the original NR data sets presented here. For 
example, all spherical harmonic modes (through $\ell = 8$) of a precessing NR waveform ($26.5$MB) 
can be compressed to $2.0$MB with a tolerance of $10^{-6}$. The same NR data hybridised
with post-Newtonian inspiral waveform modes ($85.1$MB) is compressed to just $7.1$MB for the same tolerance. 
These compressed data sets can then be evaluated fast and efficiently and resampled as desired.
\end{abstract}


\maketitle

\section{Introduction}
\label{sec:intro}
The Advanced Laser Interferometer Gravitational-Wave Observatory (LIGO)~\cite{TheLIGOScientific:2014jea} recently reported the first direct detections of gravitational waves (GWs) emitted by coalescing binary black holes (BBH)~\cite{GW150914, Abbott:2016nmj}. 
Prior to this, the coalescence of two black holes has been regarded as one of the most promising sources for ground-based GW detectors. 
The detection of GWs from such sources provides the crucial observations needed to study gravity in the strong-field dynamical regime, to test the theory of general relativity (GR), and to understand the distribution and formation mechanisms of binary black holes in galaxies, among other things.

GWs provide a unique opportunity to directly measure the source properties, such as the individual black hole masses $m_1$ and $m_2$ and the individual (dimensionless) spin angular momenta $\vec{\chi}_1$ and $\vec{\chi}_2$.
Data analysis algorithms to extract these parameters depend heavily on prior knowledge of the expected gravitational waveforms and their variation across the binary parameter space.
In the absence of globally valid analytical approximations to or solutions of the general relativistic two-body problem, much effort has been invested in developing semi-analytical waveform models that aim to represent accurately the inspiral, merger, and ringdown (IMR) waveform of black hole binaries~\cite{Ajith:2007qp, Ajith:2009bn, Pan:2009wj, Santamaria:2010yb}.
Historically, such IMR models have been constructed for non-spinning or aligned-spin BBH configurations~\cite{Ajith:2009bn, Santamaria:2010yb, Khan:2015jqa,Taracchini:2012ig, Taracchini:2013rva} but recent progress has been made for precessing cases, where the spins may be oriented arbitrarily~\cite{Hannam:2013oca,Pan:2013rra}. 
To construct IMR waveform models, numerical solutions of the non-linear Einstein field equations for BBH coalescences are necessary to calibrate the late inspiral and merger stages, as well as to determine the remnant properties (see e.g. ~\cite{Jimenez-Forteza:2016oae, Healy:2016lce, Hemberger:2013hsa}). 

Following the breakthrough in 2005~\cite{Pretorius:2005gq, Campanelli:2005dd, Baker:2005vv}, recent years have seen tremendous progress in numerically solving for the merger dynamics and associated gravitational waveforms for many different BBH configurations across the binary parameter space. This increased parameter space coverage has been particularly important for tuning IMR waveform models to a wider range of binary configurations. 
However, numerical relativity (NR) waveforms are also directly useful for GW data analysis as they provide us with the most physically realistic waveforms. In the past, the numerical injection analysis (NINJA) projects have been set up to test different data analysis pipelines~\cite{Aylott:2009ya, Aasi:2014tra} on NR waveforms. 
While the NINJA projects have been very successful in testing search and parameter estimation pipelines employed in LIGO data analysis, the injected \emph{mock} gravitational waveforms were restricted to non-spinning or aligned-spin binary systems and only contained the dominant harmonic of the GW signal. To date, however, hundreds of multi-modal NR waveforms are available that span an ever increasing range in mass ratios, spin magnitudes, spin orientations \cite{Lousto:2010ut, Mroue:2013PRL, SXSCatalog, Pekowsky:2013ska, Husa:2015iqa, Jani:2016wkt}, number of orbits~\cite{Szilagyi:2015rwa} and eccentricities. 
Among other, directly employing NR waveforms in data analysis provides the unique opportunity to independently assess the quality of semi-analytic waveform models, determine the presence of systematic modelling errors for example via Bayesian parameter estimation~\cite{Veitch:2014wba}, include important physics, like higher order modes, that is often neglected in currently available IMR model waveforms, and test the sensitivity of GW searches to these most complete waveforms available.

Due to the high computational cost of numerical simulations, most available NR waveforms are relatively short, spanning on the order of ten(s) of orbits before merger, and therefore need to be supplemented with approximate analytical inspiral waveforms such as post-Newtonian (PN) or effective-one-body (EOB) waveforms, to be also fully contained in LIGO's sensitivity band for low total masses.
This is achieved in a process known as {\it hybridisation}~\cite{Ajith:2007qp, Santamaria:2010yb, Hannam:2010ky, Boyle:2011dy, Boyle:2008ge, MacDonald:2011ne}, where NR waveforms and analytic inspiral waveforms are smoothly joined together.
For Advanced LIGO, which is expected to be sensitive to GW signals starting at 10Hz in the forthcoming years~\cite{TheLIGOScientific:2014jea}, multi-modal hybridised waveforms, which may have tens of relevant harmonics depending on the signal-to-noise ratio of a potential signal, will constitute sufficiently large data sets for analyses. 
Additionally, the previously used interface to make NR/hybrid waveforms accessible to existing data analysis pipelines, required memory intensive operations as well as intermediate storage of resampled data, which can be as big as several GB for a single GW mode. However, even if the resampling step is avoided, the raw data files for multi-modal NR/hybrid data sets may require large amounts of storage space. As an example, one of the hybrid waveforms we will be using in Sec.~\ref{sec:GW} contains 77 GW modes which requires 84MB of storage. Using the NINJA tools and a sampling frequency of 16kHz, we find that an intermediate storage of 9GB is required per GW mode, which is rather prohibitive for any analysis.
The large data sets of hybridised BBH waveforms are an obstacle to multiple-query applications in GW data analysis, such as parameter estimation studies where waveforms need to be accessed and generated many times.  In this paper, we use ideas from {\it reduced-order modelling} to efficiently and accurately compress these large data sets and present an algorithm which allows for the efficient integration of NR/hybrid waveforms into existing data analyses pipelines. 

The application of reduced-order modelling in GW physics began with the observation that the dynamics of BBHs during the inspiral phase can be dimensionally reduced suggesting that many configurations share similar qualities that vary smoothly across parameter space~\cite{Galley:2010rc}. Subsequently, reduced-order modelling techniques have been used to efficiently 
represent/compress large waveform banks~\cite{Field:2011mf, Cannon:2011xk, Herrmann:2012if, Caudill:2011kv, Blackman:2014maa} 
and to build fast and accurate surrogate models~\cite{Field:2013cfa} of merger waveforms~\cite{Field:2013cfa, Cannon:2012gq, Purrer:2014fza, Blackman:2015pia, Purrer:2015tud},
which can be used in multiple-query applications like parameter estimation studies~\cite{Canizares:2013ywa, Canizares:2014fya, Smith:2016qas} that use reduced-order
quadratures~\cite{antil2013two}.
In fact, reduced-order models are crucial in modern GW search pipelines~\cite{Cannon:2011vi} and parameter estimation studies to accelerate waveform generation and likelihood computations.

In this paper we present a fast and efficient algorithm whose advantages are twofold: Firstly, it compresses any generic large one-dimensional data set without loss of requested accuracy, and secondly provides an accurate interpolant for the uncompressed data set. 
The method we present uses a \emph{greedy algorithm}~\cite{Cormen:2001} to judiciously select only a subset of the data that allows for the reconstruction of the original values up to some requested accuracy using standard spline interpolation. We call the resulting interpolant a {\it reduced-order spline (ROS) interpolant}. Aside from providing data compression, the ROS can also be used to interpolate the original data set at new samples. As such, there are associated interpolation errors that we show how to estimate using methods from statistical learning (e.g., see~\cite{hastie2009elements}). 

We emphasise that the reduced-order spline method is generically applicable to any one-dimensional data set, but in particular to multi-modal hybrid or NR waveforms, which will be the focus of the application presented in this paper. 
Once the NR/hybrid waveforms have been compressed via the ROS method, the resulting waveforms can directly be used in LIGO GW data analysis applications via a simple interface~\cite{nrinj}, which is fully implemented in the LIGO Algorithms Library (LAL)~\cite{lal}.

The paper is organized as follows. 
In Sec.~\ref{sec:method} we describe the greedy algorithm used to generate the ROS interpolant
and elucidate some of its characteristics and features on example data generated from a test function. 
In Sec.~\ref{sec:GW} we apply the ROS interpolation algorithm to various sets of NR and hybridised gravitational waveform modes. 
We demonstrate that the data load per mode is significantly reduced. We also show that interpolating the ROS onto new samples is equally as accurate as traditional linear or cubic interpolation on the original data set, indicating that accuracy is not sacrificed here by significant data reduction and speed-up in evaluation. 

We have made publicly available a Python code called {\sc romSpline}, which is easy to use and can be downloaded from~\cite{romspline}. 
The code implements the greedy algorithm discussed in Sec.~\ref{sec:method} and includes the functionality to assess errors in the resulting reduced-order spline interpolant that are discussed in this paper. 
In addition, the repository in~\cite{romspline} contains tutorials showing how to use {\sc romSpline}.

\section{The reduced-order spline interpolation algorithm}
\label{sec:method}

Consider a set of one-dimensional, univariate data values $\vecy = \{ y_i \} _{i=1}^P$ that depend on $P$ samples, $\vecx = \{ x_i \}_{i=1}^P$. 
This data may be generated by a known function $f$ with $y_i = f (x_i)$, by numerical solutions of differential equations, by a series of measurements, etc. 
Collectively, we define $\vecd = (\vecx, \vecy)^T$ as the (original) data set, where ${}^T$ indicates transposition.

We seek to find a subset $\vecD$ of the data $\vecd$ such that a spline interpolant built from $\vecD$ will represent the dataset up to a specified tolerance $\epsilon$, as measured by the $L_\infty$-norm, which we take to be
\begin{align}
	|| \vec{\Delta} || _\infty \equiv {\rm max} \{ | \Delta_i |  \}_{i=1}^P,
\end{align}
where $\vec{\Delta}$ is the difference between $\vecy$ and the spline interpolant's prediction at the $\vecx$ samples.
The size of $\vecD$ will be relatively small if its elements are chosen in a judicious way.
To accomplish this, we use the greedy algorithm presented in Alg.~\ref{alg:greedy} and described below. We call the resulting interpolant a {\it reduced-order spline}.

{\scriptsize
\begin{algorithm}[H]
\caption{Reduced-order spline greedy algorithm}
\label{alg:greedy}
\begin{algorithmic}[1]
\State {\bf Input:}  $\vecd = (\vecx, \vecy)^T = \{ (x_i, y_i ) \}_{i=1}^P , ~ \epsilon, ~p$
\vskip 10pt
\State Set $n=p+1$  ~~(number of seeds)
\State {\bf Seed choice}:  $\vecD = (\vecX, \vecY)^T = \{ ( x_{i_k} ,  y_{i_k}  ) \}_{k=1}^{n} \subset \vecd$ ~~ (selection is arbitrary)
\While{$\sigma_n \ge \epsilon$}
\State $S_{n}^{(p)} = {\rm Spline}(\vecX, \vecY,  p)$  ~~(Build a trial spline of degree $p$ from $\vecX$ and $\vecY$)
\State $\vec{\Delta}_{n} = \{ | y_i - S_{n}^{(p)}(x_i) | \}_{i=1}^P$ 
\State $\sigma_{n} = \max ~ \vec{\Delta}_{n} $  ~~(Compute the ``greedy error'')
\State $i_{n+1} = {\rm argmax} ~ \vec{\Delta}_{n} $
\State $\vecX = \vecX \cup x_{i_{n+1}} ~{\rm and} ~ \vecY = \vecY \cup y_{i_{n+1}}$
\State $n = n+1$
\EndWhile
\State Set $N = n$
\State $S_N^{(p)} = {\rm Spline} (\vecX, \vecY, p)$
\vskip 10pt
\State {\bf Output:}  Spline interpolant $S_N^{(p)}$ and reduced data subsets $\vecD = (\vecX, \vecY)^T \subset \vecd$
\end{algorithmic}
\end{algorithm}
}

The original data $\vecd$ is the {\it training set} for the greedy algorithm, and we will often refer to $\vecd$ as such.
The algorithm starts with the full dataset $\vecd$ 
together with a tolerance $\epsilon$ up to which the reduced-order spline should recover $\vecd$. 
Additionally, we need to specify the degree $p$ of the interpolating polynomials, also known as the {\it order of the spline}.

Given these inputs the greedy algorithm proceeds as follows. First, we select $p+1$ points from the original data set, $\{ ( x_{i_k} ,  y_{i_k} ) \}_{k=1}^{p+1} \subset \vecd$, to seed the greedy algorithm. These points may be chosen arbitrarily and, in general, the algorithm described in Alg.~\ref{alg:greedy} allows for arbitrary seeds as does our publicly available code implementation \textsc{romspline}~\cite{romspline}. In \textsc{romspline} the default seeds consist of the first and last data points together with $p-1$ (nearly) equally-spaced points in between. 
We will refer to $\vecD = (\vecX, \vecY)^T$ as the {\it reduced data} set. 
At this stage, the reduced data comprises only the seed elements.

Second, we build a first (trial) spline of order $p$ from the seed data. 
We do not use a smoothing or regularisation factor when building the spline. 
Instead, we force the spline to reproduce the reduced data at machine precision. 
We have found through numerical experimentation that enabling smoothing tends to prevent the algorithm from reaching the requested tolerance $\epsilon$, especially for $\epsilon \lesssim 0.1$. 
In addition, we are not concerned with the particular type of spline used (Hermite spline, Bezier spline, etc.).

Third, we compute the absolute value of the point-wise difference between the training data values $\vecy$ and the trial spline's predictions at the training samples, $\vecx$.
The element $i_{n+1}$ of $\vecd$ with the largest such difference is selected and the corresponding pair $(x_{i_{n+1}}, y_{i_{n+1}})$ is added to the reduced data $\vecD$. 
Upon the following iteration in the greedy algorithm, the next trial spline will represent that newly added value $y_{i_{n+1}}$ to within numerical round-off precision as we are not implementing smoothing splines.
This process is repeated until the maximum point-wise error is below the specified tolerance $\epsilon$. 

Lastly, from the reduced data $\vecD$ generated by the greedy algorithm we build a final spline interpolant, which is the reduced-order spline, $S_N^{(p)}(x)$.
By construction, $\vecD$ is reproduced by $S_N^{(p)}$ to within machine precision while all other elements in $\vecd$ are recovered to within an error less than the specified tolerance $\epsilon$, as measured by the $L_\infty$-norm.

We remark that Alg.~\ref{alg:greedy} is hierarchical (or nested) in the sense that the elements
selected by the greedy algorithm in the $n^{\rm th}$ step (i.e., $(x_{i_{n+1}}, y_{i_{n+1}})$) are {\it added} to the current reduced data set. 
Therefore, the accuracy of the ROS can be improved by adding more points 
(as selected by the greedy algorithm) to $\vecD$. If less accuracy
is needed then the appropriate number of latest entries is excluded from $\vecD$ until the desired tolerance is reached. 
The greedy algorithm can be used with interpolants other than splines. In addition, Alg.~\ref{alg:greedy} can be generalised to higher dimensions with splines, for example, on regularly spaced data.

In the following subsections, we perform some numerical experiments to gauge the behaviour 
and properties of the greedy algorithm. To provide context, we sample the function
\begin{equation}
	f(x) = 100 \left[ (1+x) \sin \big( 5 (x-0.2)^2 \big) + \exp \left(-\frac{ (x-0.5)^2 }{  0.02 }  \right) \sin( 100 x) \right]
\label{eq:function}
\end{equation}
at values $\{x_i\}_{i=1}^P$ so that $y_i = f(x_i)$.
This function describes a chirping sinusoid with a linearly growing amplitude and a Gaussian modulated by a high frequency sinusoid. 
The top panel in Fig.~\ref{fig:test_data} shows a plot of the function in Eq.~\eqref{eq:function} for 
$x \in [-1, 1]$ at $4001$ uniformly spaced samples, $\vecx$.
We choose to work with a model function in this section to assess the true errors in using the ROS and 
to determine the reliability of our interpolation uncertainty estimates (see Sec.~\ref{sec:cvTest}).

\subsection{Properties of the greedy algorithm}
\label{sec:greedyprops}

The function in Eq.~\eqref{eq:function} maps samples $\vecx$ to values $\vecy$ so that our training set is 
$\vecd =  \{ ( x_i, y_i = f(x_i) ) \}_{i=1}^P$.
We next apply the greedy algorithm of Alg.~\ref{alg:greedy} to $\vecd$.
We choose a default tolerance of $\epsilon = 10^{-6}$ and use spline polynomials of degree $p=5$. 
The output of Alg.~\ref{alg:greedy} includes a subset of $\vecd$ that constitutes the reduced data set 
$\vecD = \{ ( X_j, Y_j ) \}_{j = 1}^N$ 
where $\vecX = \{ x_{i_k} \} _{k=1}^N$ and $\vecY = \{ y_{i_k } \} _{k=1}^N$.

For this test case, the greedy algorithm produced a reduced data set containing $441$ elements in $\vecX$ and $\vecY$.
Therefore, the resulting ROS interpolant $S_{441}^{(5)}(x)$ requires only these $441$ points to reconstruct the original data values $\vecy$ at the samples $\vecx$ to a tolerance of $10^{-6}$ in the $L_\infty$-norm.
These numbers lead to a compression factor of $4001 / 441  \approx 9.1$.

\begin{figure}
\begin{center}
\includegraphics[width=0.49\textwidth]{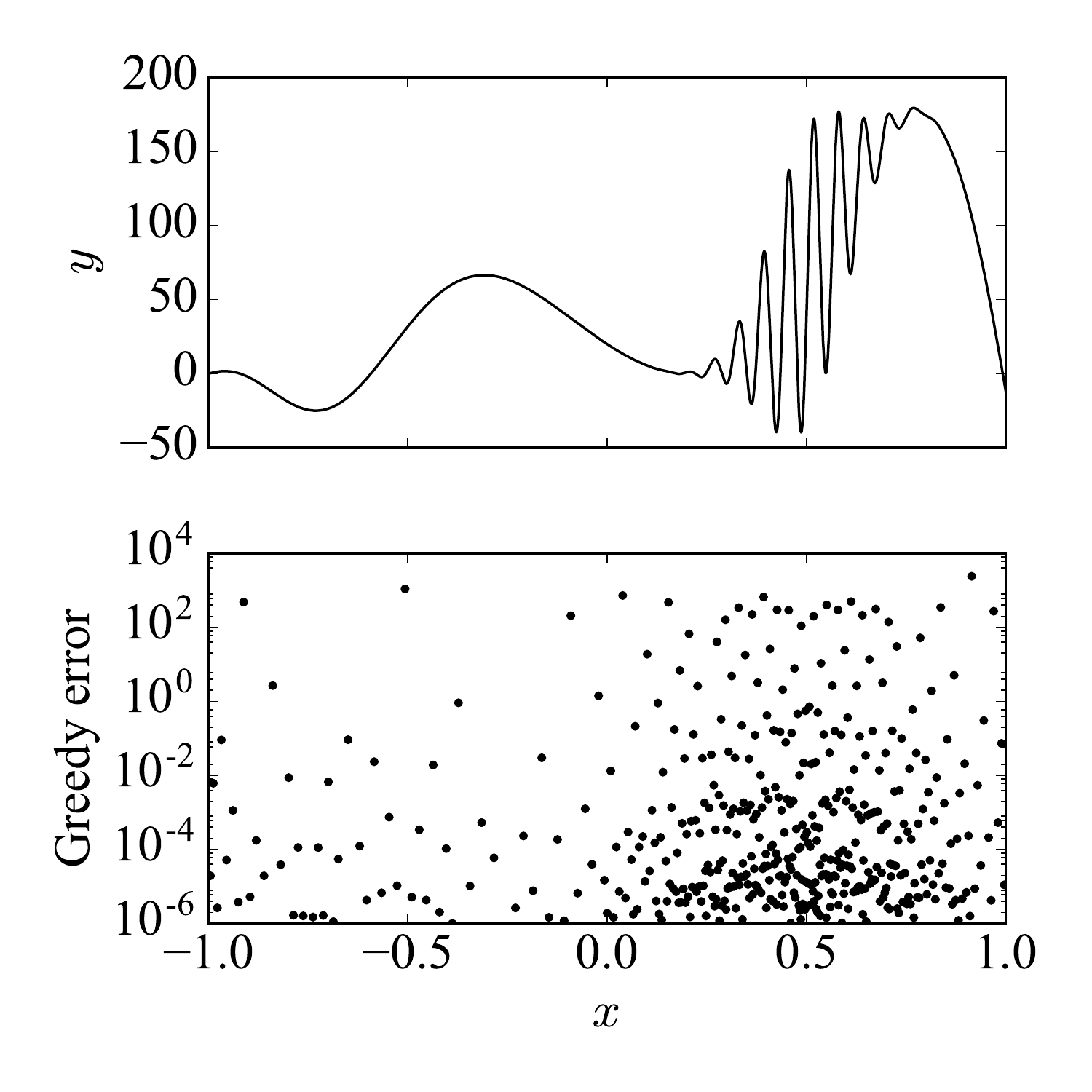}
\includegraphics[width=0.49\textwidth]{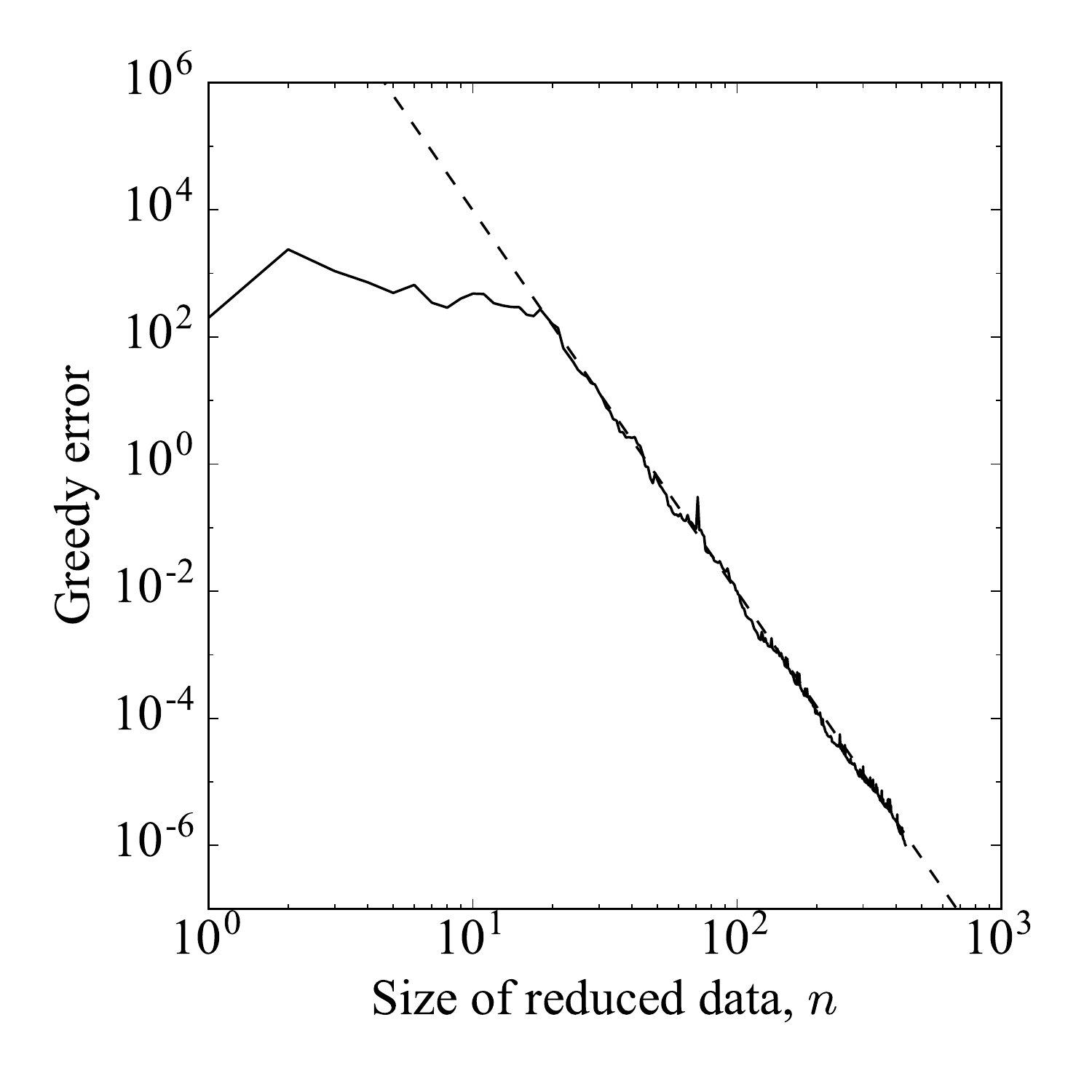}
\caption{{\bf Top left}: The test data used to provide context for revealing the performance and properties 
		of the reduced-order spline interpolant greedy algorithm.
		{\bf Bottom left}:  The distribution of the corresponding (reduced) data samples selected by the greedy algorithm       shown with the corresponding greedy errors on the vertical axis.
		{\bf Right}: The greedy error for building a reduced-order spline interpolant of degree $p=5$ 
		with a specified tolerance of $\epsilon = 10^{-6}$ in the $L_\infty$-norm.
		After about $20$ steps, the greedy error decays as $n^{-(p+1)}$. 
		The dashed line is a fit to $A / n^{5+1}$ for $n \ge 20$ with $A \approx 10^{10}$.
		 }
 \label{fig:test_data}
\end{center}
\end{figure}

The (nonuniform) distribution of the $\vecX$ elements is shown in the bottom left panel of Fig.~\ref{fig:test_data}.
Notice that more points are selected around the regions that exhibit the strongest variation with $x$.
This is a common feature of greedy algorithms as stronger features in the training set tend to require more points to resolve with sufficient accuracy. 
As a consequence, we cannot use $441$ approximately equidistant samples to represent the original data set with a spline interpolant of the same degree to a tolerance of $\epsilon = 10^{-6}$.
For comparison, a spline built using $441$ equally spaced samples is found to have a maximum absolute error over $\vecd$ of $6.7 \times 10^{-5}$. We would require $862$ equally spaced samples (i.e., nearly twice the size, $441$, of the reduced data set) to generate a spline satisfying $\epsilon \le10^{-6}$.

At each step $n$ in the greedy algorithm, we compute the maximum absolute value of the point-wise difference between the training data values $\vecy$ and the predictions of the current ROS interpolant $S_n^{(p)}$
at the training samples $\vecx$,
\begin{align}
	\sigma_n = \max \{ | y_i - S_n^{(p)} (x_i) | \} _{i=1}^P \, .
\end{align}
We call the set of these values, $\{ \sigma_n \}_{n=1}^N$, the {\it greedy errors} of the ROS interpolant $S_N^{(p)}(x)$.
The right panel of Fig.~\ref{fig:test_data} shows how the greedy errors change with $n$, the size of the $n^{\rm th}$ trial reduced data set.
After about $20$ steps (i.e., when $\vecD$ contains $20+p+1$ points from $\vecd$)
the greedy error converges to the requested tolerance of $10^{-6}$ as an inverse power of $n$. 
The dashed line is a fit to the $n \ge 20$ greedy errors with the function $A / n^6$ with $A \approx 10^{10}$. 

Fig.~\ref{fig:greedy_deg} shows the greedy errors versus the reduced data size $n$ for several polynomial degrees $p$, with everything else held the same.
We observe that the greedy error decays with $n$ as an inverse power dependent on the degree of the polynomial used for the ROS.
The dashed lines are fits of the greedy errors to the function $A_p / n^{p+1}$ for $p=1, 2, 3, 4, 5$. 
The dotted vertical line indicates the size of the original data set, $\vecd$. 
We find  that the decay changes to being approximately exponential with $n$ once $\vecD$ contains about one quarter of the points in $\vecd$. 
For larger $p$, this transition occurs at higher tolerances than shown in the plot.

\vskip0.25in
\noindent
\begin{minipage}{\textwidth}
\begin{minipage}[b]{0.48\textwidth}
	\centering
	\includegraphics[width=\textwidth]{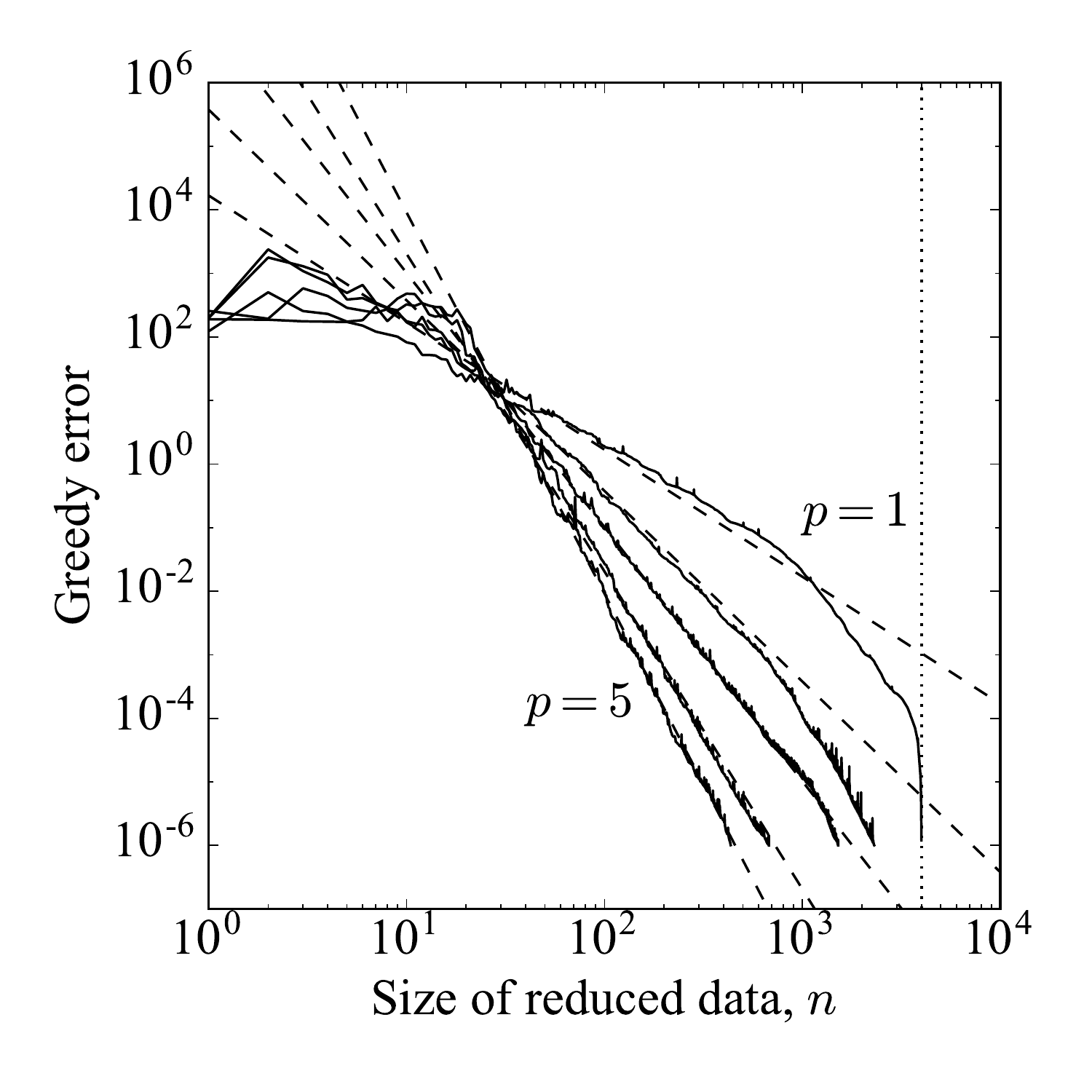}
	\captionof{figure}{
		The greedy error for building a reduced-order spline interpolant of degree $p=1,2,3,4,5$ 
		with a specified tolerance of $\epsilon = 10^{-6}$ in the $L_\infty$ norm.
		The dashed lines are fits to $A / n^{p+1}$.
		The dotted vertical line is the number of points in the original data.
		}
	\label{fig:greedy_deg}
\end{minipage}
\hfill
\begin{minipage}[b]{0.48\textwidth}
	\centering
	\begin{tabular}{c c c}
		\hline\hline
		Degree, $p$ & Size, $N$ & Compression \\
		\hline
		1 & 3,994 & 1.002 \\
		2 & 2,308 & 1.734 \\
		3 & 1,520 & 2.632 \\
		4 & 683 & 5.858 \\
		5 & 441 & 9.073  \\
		\hline\hline
	\end{tabular}
	\vskip0.6in
	\captionof{table}{
		The size of the reduced data set $N$ and compression $C$ versus 
		the degree of the interpolating polynomial $p$ used in building the reduced-order spline.
		The tolerance on the greedy error is the same in all cases, $\varepsilon = 10^{-6}$.
		}
	\label{tab:degrees}
\end{minipage}
\end{minipage}
\vskip0.25in

Fig.~\ref{fig:greedy_deg} also indicates that the size $N$ of $\vecD$ tends to be inversely related to the degree $p$ of the ROS polynomial. 
The dependence of $N$ on $p$ is tabulated in Table~\ref{tab:degrees}. 
Notice that $\vecD$ for the case with $p=1$ contains all but seven points from the original data set. 
In general, low-order interpolating polynomials are not useful for data compression because of their lower degree of flexibility.
Therefore, throughout the remainder of this paper we will use $p=5$ unless otherwise noted.

\subsection{Effects of choosing different seed data}
\label{sec:seed1}

The output of Alg.~\ref{alg:greedy} depends on the subset of data used to seed the greedy algorithm in addition to the tolerance $\epsilon$ and the spline degree $p$.
Our next numerical experiment explores how the size of the reduced data set $\vecD$ depends on the seed used to initialise the greedy algorithm. 

\begin{figure}
	\includegraphics[width=0.5\columnwidth]{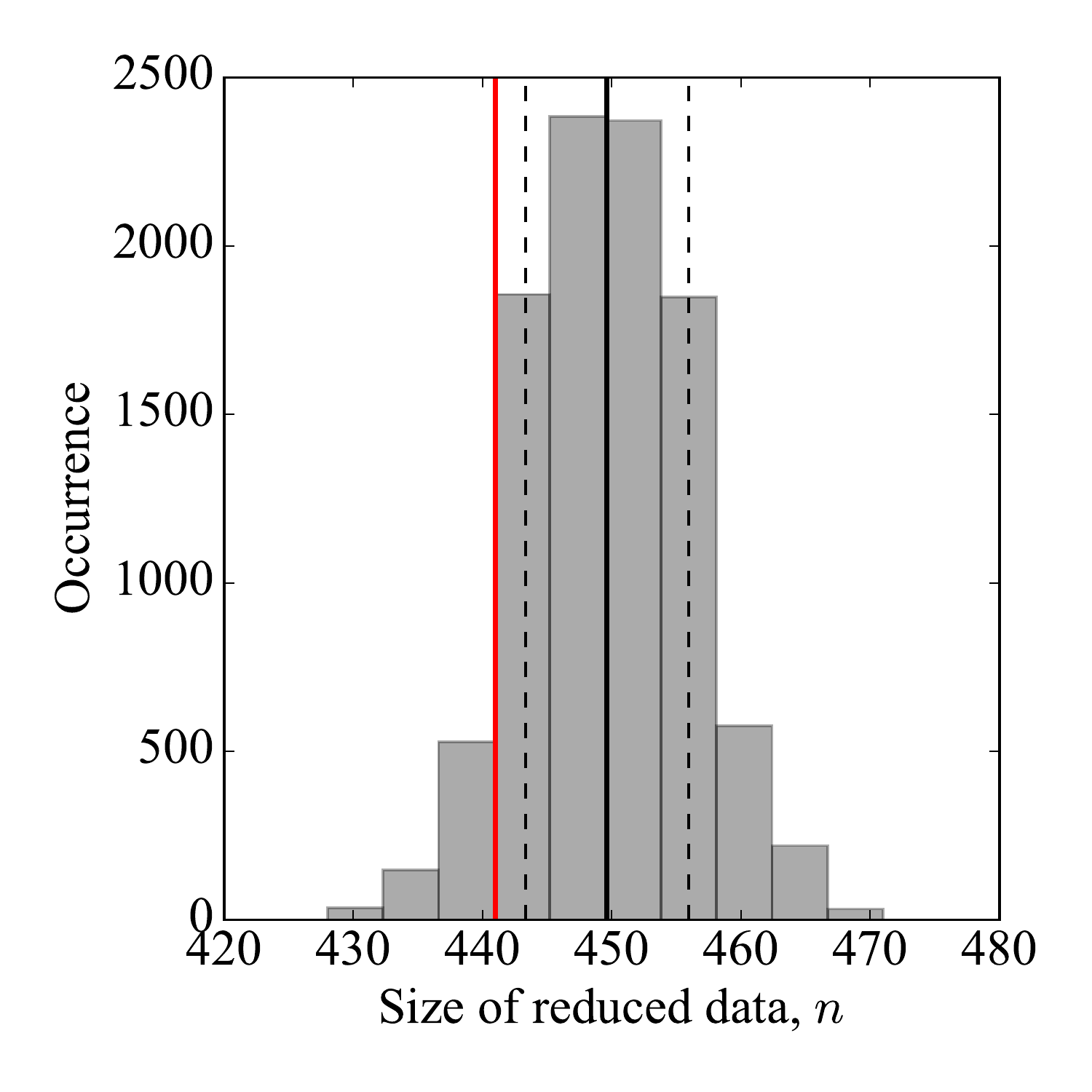}
	\hspace{-0.15in}
	\includegraphics[width=0.5\columnwidth]{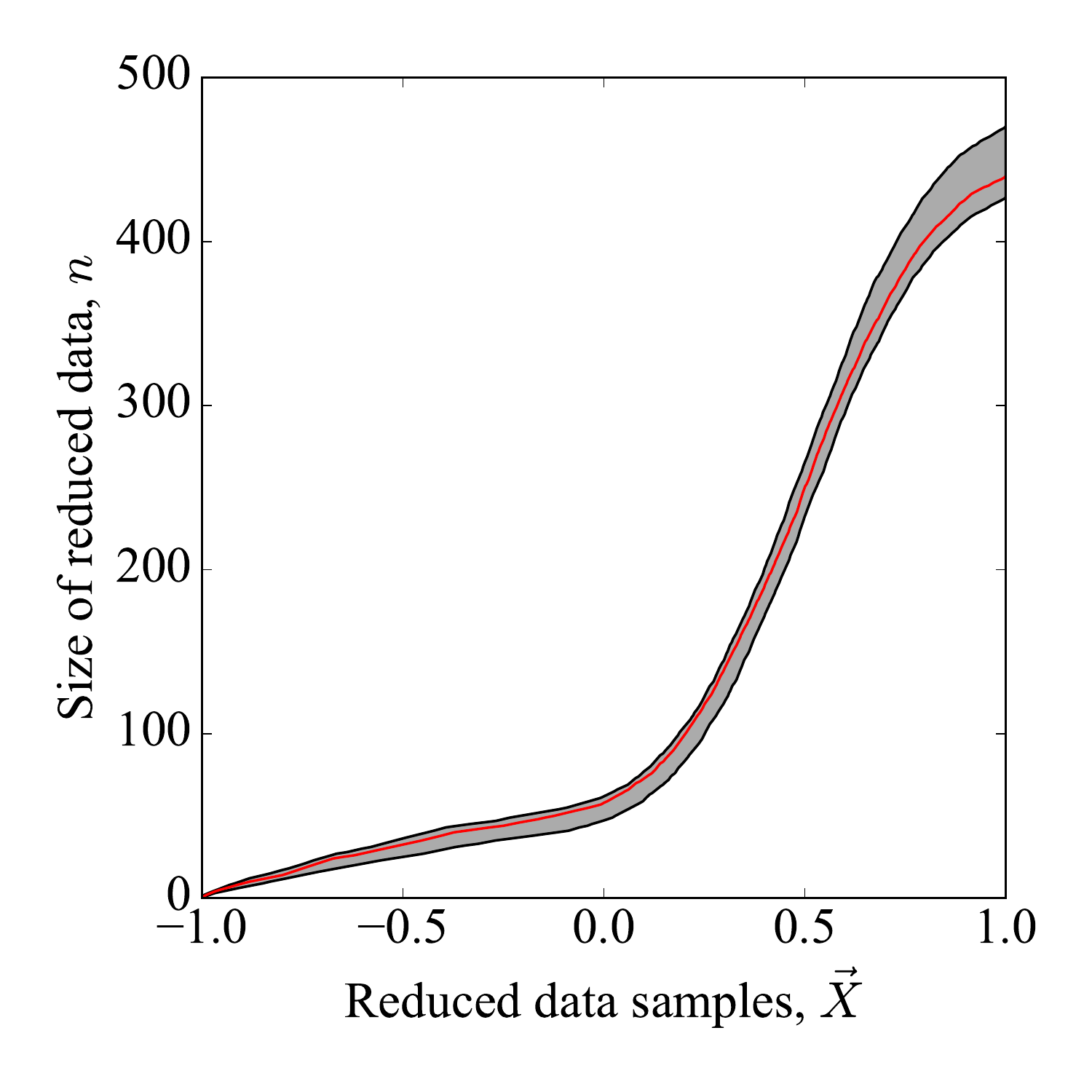}
	\caption{
		{\bf Left}:
		The distribution of the reduced data sizes for $10,\!000$ randomly selected sets of seeds.
		The mean (solid black line) and sample standard deviation (dashed black lines) 
		of the distribution are $449.6$ and $6.3$, respectively.
		The smallest (largest) size in this sample has $428$ ($471$) elements. 
		The red line is the size of the reduced data for the fiducial reduced-order spline
		in Sec.~\ref{sec:greedyprops}.
		{\bf Right}:
		The size of the reduced data as a function of the (sorted) reduced 
		samples $\vecX$ selected during the greedy algorithm for 
		each of the $10,\!000$ random seeds of the left panel.
		The reduced samples of the fiducial reduced-order spline are shown in red.
		While the samples selected for each of the $10,\!000$ cases are different, 
		the narrowness of the shaded region indicates that nearly the same samples
		are selected.
		}
	\label{fig:seeds}
\end{figure}

We randomly select $10,\!000$ sets of seeds and generate a ROS for each. Each seed set contains $p+1$ points from the original data set $\vecd$.
We keep $p=5$ and $\epsilon=10^{-6}$ fixed.
The left panel in Fig.~\ref{fig:seeds} shows how the sizes $N$ of those reduced data sets are distributed.
We observe that the sizes are approximately normally distributed with a mean of $449.6$ 
(solid black line) and a sample standard deviation of $6.3$ (flanking dashed black lines).
The red vertical line indicates the reduced data size that corresponds to the seed used for building the fiducial ROS in the previous section.

The smallest reduced data set has $428$ elements while the largest has $471$.
These values correspond to approximately a $4.8\%$ and $4.7\%$ spread from the mean value, respectively. 
Therefore, while a particular seed might not give rise to the smallest possible number of points needed to represent the original data with the ROS, it will have {\it nearly} the smallest number of points (typically within a few percent). 
Every seed generates an optimally reduced data set {\it for that seed} but it is not likely to be the smallest possible such subset.
For this reason, the greedy algorithm in Alg.~\ref{alg:greedy} can be said to yield a {\it nearly} optimal reduction of $\vecd$ given $p$, $\epsilon$, and $p+1$ seed elements.

The right panel in Fig.~\ref{fig:seeds} shows how the reduced data samples $\vecX$ vary (shaded region) for 
each of the $10,\!000$ seed sets. The fact that the shaded region is fairly narrow over the interval of $\vecx$ suggests that the distribution of $\vecX$ is robust to changes in the seed that initialises the greedy algorithm. Notice that the gradient of the shaded region steepens where the data has more structure (see Fig.~\ref{fig:test_data}) and is relatively flat where the data is in the low frequency region.

\subsection{Convergence of reduced-order spline interpolation with training set density}

We next discuss the convergence of the greedy algorithm by decimating the original data set and studying the accuracy of the resulting reduced order splines on $\vecd$.
The algorithm converges if subsequent decimations leave the size of the corresponding reduced data sets approximately unchanged.

\begin{figure}[]
	\includegraphics[width=0.5\columnwidth]{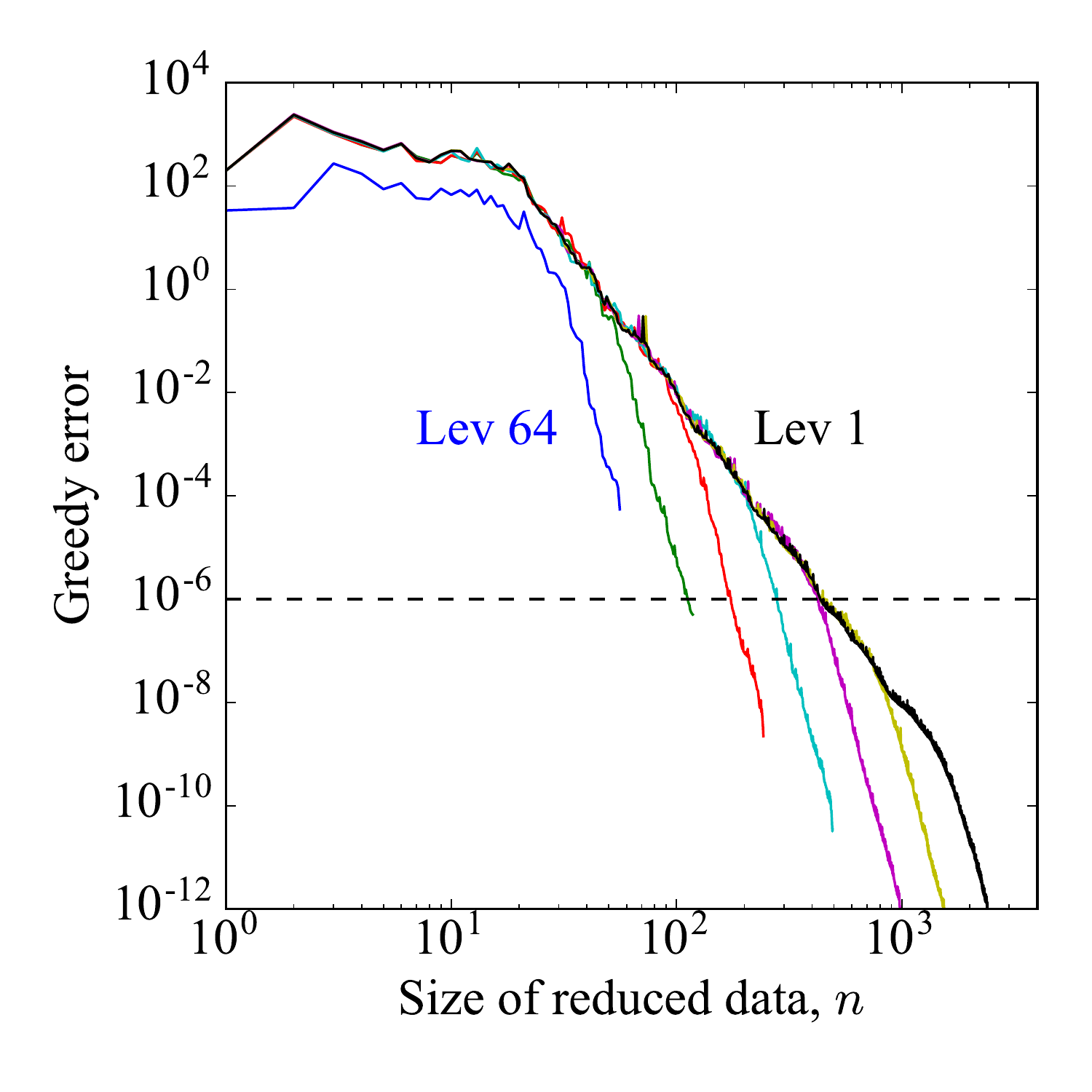}
	\caption{
	Convergence of the reduced-order spline with the number of data samples
	used to train the greedy algorithm for this example. The original data set is decimated 
	by a factor of $2^n$ (denoted ``Lev $2^n$'') for $n=0,1,2,\ldots, 6$
	 and the resulting greedy errors
	are plotted. 
	For a tolerance of $10^{-6}$ (dashed line) convergence is achieved at Lev $4$.
	}
	\label{fig:convergence}
\end{figure}

Decimation can be regarded as changing the level (i.e., ``Lev'') of resolution of the data.
Fig.~\ref{fig:convergence} shows the greedy errors when every $2^n$th point
is used from the test data set $\vecd$ to construct a trial ROS. 
We take $n = 0, 1, \ldots, 6$ and refer to the corresponding trial as ``Lev $2^n$''. 
For the example data set, convergence in the greedy errors is
reached by Lev 4 for a tolerance of $10^{-6}$ (dashed line). 
Hence, the Lev 4 ROS reproduces $\vecy$ at the original samples $\vecx$
to the same accuracy as the Lev 2 and Lev 1 splines. 
Being in the convergent regime, the reduced data associated with these three ROS's have approximately the same size.
However, convergence may not be reached for smaller $\epsilon$. 
For example, the greedy error is not convergent for $\epsilon = 10^{-12}$ (see Fig.~\ref{fig:convergence}) despite the fact that, {\it by construction}, each ROS represents the original data to this level of accuracy.
The convergence, or lack thereof, has consequences for estimating the ROS errors in predicting values at {\it new} samples not in $\vecd$.

The relationship between the convergence of the ROS and 
$\epsilon$ is a property of $\vecd$. 
For example, the ROS is convergent
for $\epsilon = 10^{-6}$ but not $\epsilon = 10^{-12}$
because the original data does not sufficiently resolve the function in Eq.~\eqref{eq:function}. 
A lack of convergence for the requested tolerance indicates that not enough data is given
to build a reliable reduced-order spline or, equivalently, that the given data produces a convergent 
interpolant for a lower tolerance.

\subsection{Cross-validation to estimate the uncertainty in predicting new values}
\label{sec:cvTest}

Most interpolation applications involve generating new data values at samples not used to build the interpolant.
In this section, we quantify the uncertainty associated with using the ROS interpolant to {\it predict} a value not contained in $\vecd$.

We use the method of {\it $K$-fold cross-validation} \cite{hastie2009elements} to provide an error or uncertainty estimate on the ROS prediction of new values. 
$K$-fold cross-validation (CV) proceeds as follows. 
Partition the original data set $\vecd$ into $K$ non-overlapping subsets such that each contains $\approx 1/K$ data points\footnote{The $K$ subsets can have different sizes
but we choose them to have as nearly equal numbers of elements as possible.}. 
The elements in each partition are randomly selected from $\vecd$ and the union of these $K$ subsets is $\vecd$.
Next, we select the first subset as a {\it validation} set and train the greedy algorithm in Alg.~\ref{alg:greedy} on the remaining $K-1$ subsets to build a trial ROS interpolant.
Then, we compute the $L_\infty$-norm error of the trial ROS on the validation set. This procedure characterises the error in using the trial ROS to predict the data values in the validation set. 
Finally, this step is repeated so that each of the $K$ subsets has been used as a validation set.
This results in $K$ {\it validation errors} that can be used to estimate the uncertainty that the full ROS 
(i.e., the interpolant $S_N^{(p)}$ built from all of $\vecd$) has for predicting values at new samples.

\begin{figure}[h]
	\includegraphics[width=0.5\columnwidth]{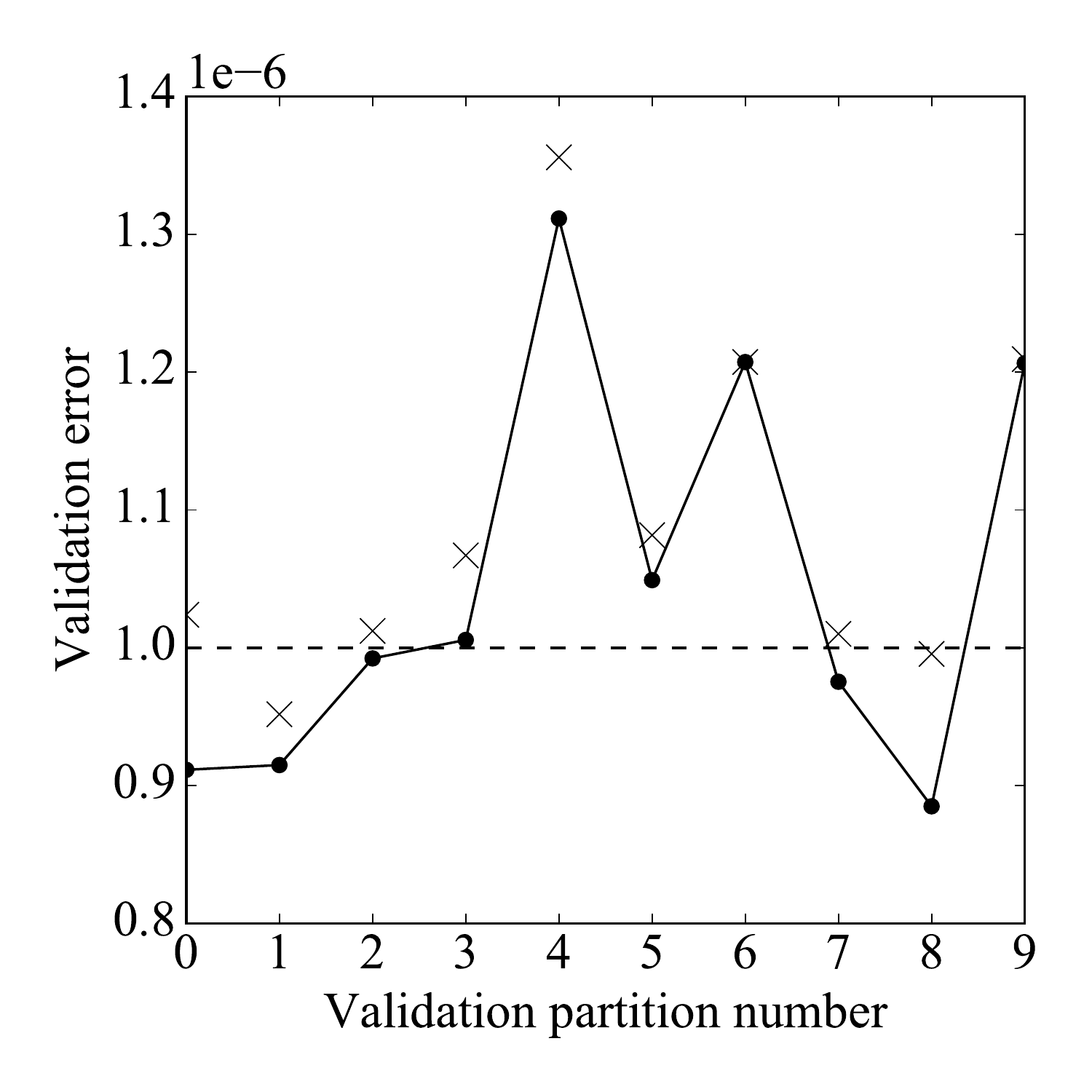}  
	\includegraphics[width=0.5\columnwidth]{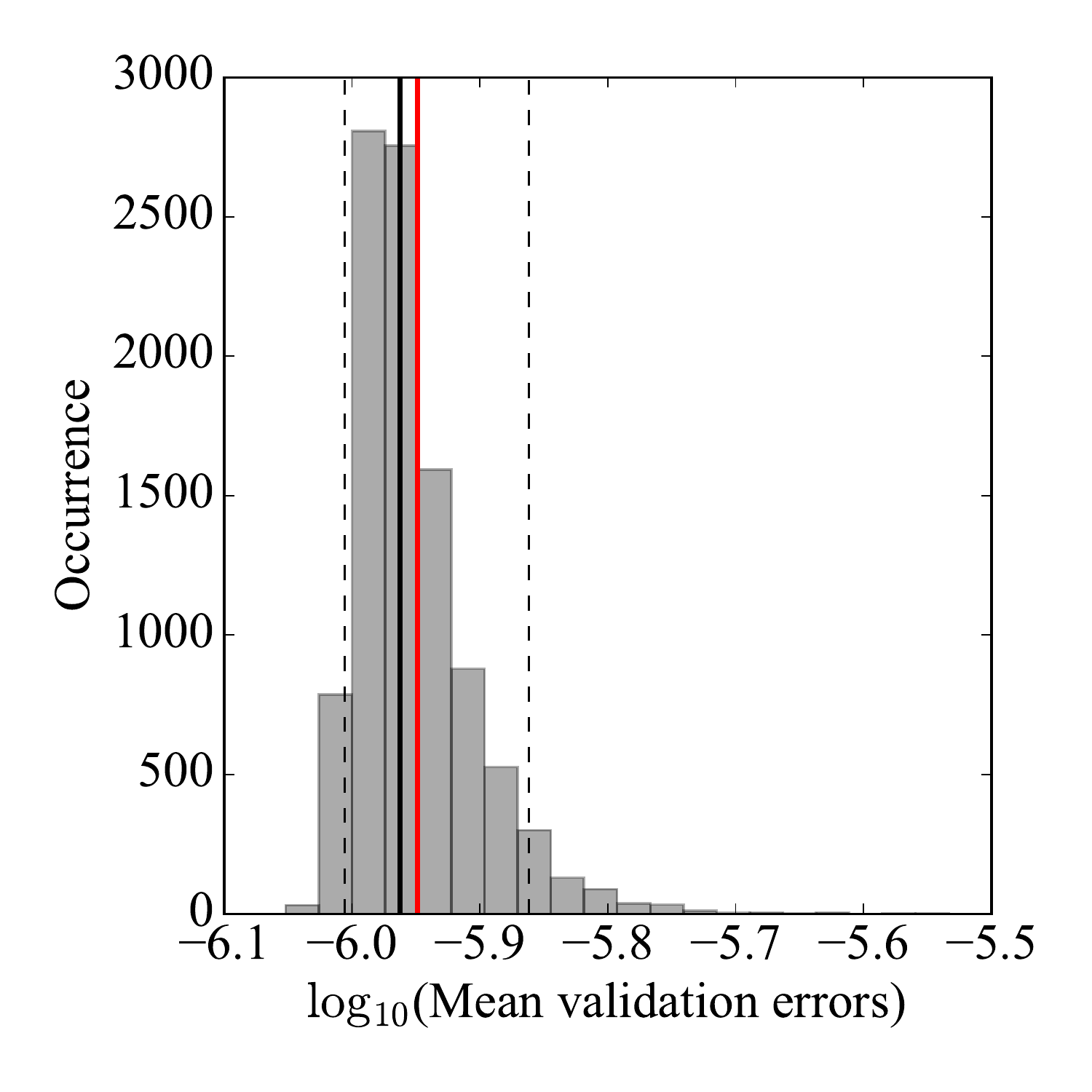}
	\caption{
	{\bf Left:}
	The validation errors of a $K$-fold cross-validation study 
	for one realization of a random distribution of the original test 
	data $\vecd$ into $K=10$ subsets (dots).
	The mean of these validation errors is $1.05 \times 10^{-6}$.
	Also shown are the maximum absolute errors (crosses) between the trial splines made from each training partition and the function in~\eqref{eq:function} at a very dense set of samples.
	{\bf Right:} 
	The distribution of mean validation errors associated with
	$10,000$ $K$-fold cross-validation studies.
	Also shown are the mean (red), median (solid black), 
	$5$th and $95$th percentiles (dashed black) of the mean errors.
	The largest error is $1.58 \times 10^{-5}$ but is highly unlikely to be realised in practice.
	}
	\label{fig:kfold}
\end{figure}

The left panel of Fig.~\ref{fig:kfold} shows an outcome of $K$-fold cross-validation for $K=10$ subsets. 
In many applications, $K=10$ is a good choice because $10\%$ of the original data is reserved for validating the trial ROS that was trained on the other $90\%$ of the values but this may depend on the particular data set in consideration~\cite{breiman1992submodel, kohavi1995study, hastie2009elements}.
The validation errors are plotted for each validation set in the left panel of Fig.~\ref{fig:kfold} as dots.
The crosses indicate the maximum absolute errors between the trial splines made from each training partition and the function in~\eqref{eq:function} at a very dense set of samples. 
As such, the crosses indicate the ``true'' interpolation error that would be found using the trial splines.
Notice that the errors shown by the crosses are always larger than those of the dots but are always comparable.
The dashed horizontal line at $10^{-6}$ is the specified tolerance for building the fiducial ROS.
We see that the validation errors tend to be comparable to, if not slightly larger than, the specified tolerance. The mean of these validation errors for this particular realisation is $1.05 \times 10^{-6}$
while the mean of the ``true'' errors associated with the crosses is $1.09 \times 10^{-6}$. 
For comparison, the ``true'' interpolation error of the fiducial spline is $1.01 \times 10^{-6}$. This is determined by evaluating the function in~\eqref{eq:function} at a very dense set of samples and finding the largest absolute difference when compared to the fiducial ROS predictions at those same points.

An accurate estimate of the interpolation uncertainty can be found by repeating the $K$-fold 
cross-validation study many times, once for each realisation of a random distribution of $\vecd$ into the $K$ subsets. 
Since each random distribution into the $K$ subsets is independent from each other then we can apply standard statistics to the resulting set of validation errors.
We call this method \emph{Monte Carlo $K$-fold cross-validation}.

We performed $K$-fold cross validation $10,000$ times on our test data $\vecd$ to form a distribution of independent mean validation errors.
The distribution of the mean validation errors for each study
 is plotted in the histogram in the right panel of Fig.~\ref{fig:kfold}.
The mean of the mean errors (red) is $1.13 \times 10^{-6}$ 
and the median (solid black) is $1.09 \times 10^{-6}$. 
The $5$th and $95$th percentiles of the maximum errors (dashed black) are $9.87 \times 10^{-7}$ and $1.38 \times 10^{-6}$, respectively.
The largest error over all $10,000$ studies is $1.58 \times 10^{-5}$
that, despite being an order of magnitude more than the specified tolerance of $10^{-6}$, 
is still a small number and highly unlikely to be realised from the plotted distribution. In addition, the largest error is an overly conservative
upper bound on the fiducial ROS's accuracy of predicting values at new samples.

For the sake of comparison, we also sample the function in Eq.~\eqref{eq:function} at $100\times 4001$ values (i.e., $100$ times more samples than the size of the original data set) randomly and uniformly drawn from $[-1,1]$.
We then compute the absolute differences between the ROS predictions and the actual function values and find that the largest interpolation error  is $1.01 \times 10^{-6}$. 
Comparing this to the mean ($1.13 \times 10^{-6}$) and median ($1.09 \times 10^{-6}$) of the mean cross-validation errors  computed above shows that the Monte Carlo $K$-fold cross-validation gives a reasonable (upper bound) estimate on the ROS interpolation error.

In situations where $10,000$ trials would be expensive to compute (e.g., because the original data set is very large and/or $\epsilon$ is very small), one may find a good
estimate, typically within about a percent, of the distribution of the Monte Carlo $K$-fold cross-validation errors with an ensemble of $100$ trials. This is because the uncertainty in the mean of the mean validation errors is inversely proportional to the square root of the number of (independent) trials.

We end this subsection with a comment about boundary points and our implementation of $K$-fold cross-validation in our code {\sc romSpline}~\cite{romspline}, which we used to obtain the results here. 
When one of the $K$ subsets is chosen as a validation set there is a choice about whether or not to include the endpoints of the original dataset to seed the greedy algorithm on the remaining data used for training. 
We have chosen to use the default seed choices mentioned earlier to the remaining training set under consideration. 
In particular, an endpoint (or both) of the original dataset may lie in a validation set so that one is extrapolating the corresponding trial ROS to the boundary point(s). 
As a result, one could worry that there might be a large (extrapolation) error incurred because of this convention. 
However, as $K$-fold cross-validation uses {\rm all} the $K$ subsets for both validation and training and because the {\it mean} of the largest absolute errors for each validation set is recorded as the CV error then any such bias from such an extrapolation is ``diluted'' by the other $K-1$ validation errors. 
Furthermore, such extrapolation errors on a particular validation subset are comparable with interpolation errors and we have yet to see a case where this does not occur, let alone egregiously so.
However, it is straightforward to always choose to include the boundary points of the original dataset in the seed choice.

\subsection{Reduced-order spline interpolation for non-smooth data}
\label{sec:noisy}

In some applications, the data may not correspond to smoothly varying functions of the samples
but may instead exhibit small-amplitude stochastic, high-frequency or generally non-smooth features.
Such a scenario may be realised by data taken from experimental measurements or observations.
Of course, if the data exhibit large-amplitude stochastic or high-frequency features then
interpolating the data for compression or prediction is not necessarily appropriate.

\begin{figure}[h]
	\includegraphics[width=\columnwidth]{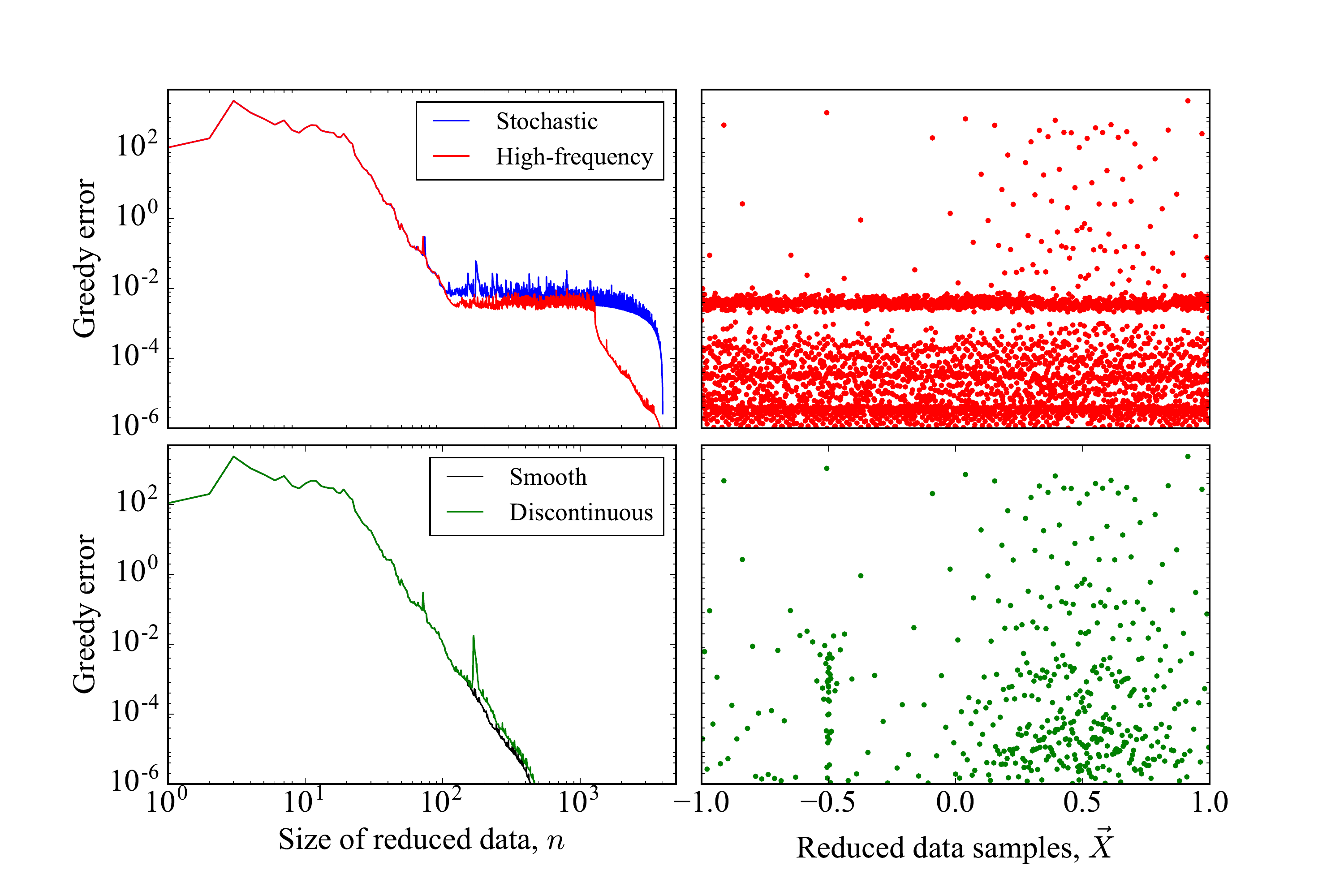}
	\caption{
	{\bf Upper left}: The panel shows the greedy errors resulting from Alg.~\ref{alg:greedy} when building
	reduced-order splines for data with low-amplitude ($\approx 10^{-3}$) stochastic (blue) and
	high-frequency (red) features.
	{\bf Upper right}: The panel shows the greedy errors versus the corresponding subsamples $\vec{X}$ selected 
	by the greedy algorithm for the high-frequency data. The high-frequency features are 
	observed as a stripe for errors near $\sim 10^{-3}$.
	{\bf Lower left}: Greedy errors for the smooth data (black) and when including a low-amplitude ($10^{-3}$) discontinuity (green). Note the spike at $n\approx 200$ when the discontinuity is encountered by the greedy algorithm.
	{\bf Lower right}: Greedy errors versus the corresponding selected subsamples $\vec{X}$ for the data with a discontinuity. Note the increased density of points near $x=-0.5$ where the discontinuity occurs in the dataset.
	}
	\label{fig:noisy}
\end{figure}

We consider three example cases: stochastic noise, high-frequency oscillations, and a $C^0$-discontinuity in the data. We include stochastic features to the smooth data in Eq.~\eqref{eq:function} by adding small-amplitude ($10^{-3}$ for these studies), normally-distributed noise to each $y_i$. 
We include high-frequency, or ``ultraviolet'' (UV), features by adding the deterministic function
\begin{align}
	f_{\rm UV} (x) = 10^{-3} \sin \left( \frac{4001}{0.01} x \right)
	\label{eq:uv}
\end{align}
to Eq.~\eqref{eq:function}. Fig.~\ref{fig:noisy} shows the greedy errors resulting from applying the Alg.~\ref{alg:greedy} separately to both the stochastic (blue) and UV (red) test data.
The defaults of $\epsilon = 10^{-6}$ and degree $p=5$ are used in building both ROS interpolants.
In both cases, a plateau in the error is observed as the greedy algorithm attempts 
to resolve the strongly varying and/or non-smooth features in the original data sets. 
The greedy algorithm is unable to resolve the noisy features in the stochastic data
until almost all original sample points have been used to build the ROS interpolant, resulting in
virtually no compression.
However, the expected polynomial convergence rate of $\propto n^{-(p+1)}$ is observed to continue
after the UV features have been resolved for the high-frequency case.

Lastly, we include a discontinuity to the smooth data in Eq.~\eqref{eq:function} by adding the small-amplitude $C^0$ function
\begin{align}
	f_{\rm C^0} (x) = 10^{-3} \theta (x+0.5)
\end{align}
where $\theta$ is the Heaviside function and the discontinuity occurs at $x=-0.5$, which is chosen to be somewhat removed from the rapidly varying features near $x = 0.5$ for visualisation purposes.
The bottom two panels in Fig.~\ref{fig:noisy} show the greedy errors and the corresponding distribution of the selected subsamples $\vecX$. Notice that at $n\approx 200$ the greedy errors are near $10^{-3}$ and begin to resolve the discontinuity in the data at $x=-0.5$. Resolving this discontinuity initially incurs relatively large errors because smooth polynomials are being used, which accounts for the spike seen in the lower left panel. As such, this requires including more data subsamples as can be seen in the lower right panel where there is an increased density of points near $x=-0.5$. The lower right panel should also be compared to the lower left panel in Fig.~\ref{fig:test_data} for the distribution of the selected subsamples for the original smooth test dataset.

\section{Application of reduced-order spline interpolation to gravitational waveforms}
\label{sec:GW}

In the previous section, we have introduced and characterised the general properties of the reduced-order spline
greedy algorithm given in Alg.~\ref{alg:greedy} and implemented in {\sc romSpline}~\cite{romspline}. 
Using example test data generated by a known function, 
which can be resampled as desired, we have assessed the uncertainties in predicting new data. 
Having understood the properties and prediction uncertainties of reduced-order spline 
interpolation, we are now in a position to apply this method to various harmonic modes of gravitational waveforms from coalescing black hole binaries produced by numerical relativity simulations. Specifically, in this section, we build and assess the efficacy of reduced-order splines when applied to NR/hybrid waveforms and its usage in GW data analysis, where fast evaluation and resampling are crucial in many analyses.

\subsection{Numerical relativity and hybrid waveforms}
\label{sec:NR}

\begin{figure}
\includegraphics[width=\columnwidth]{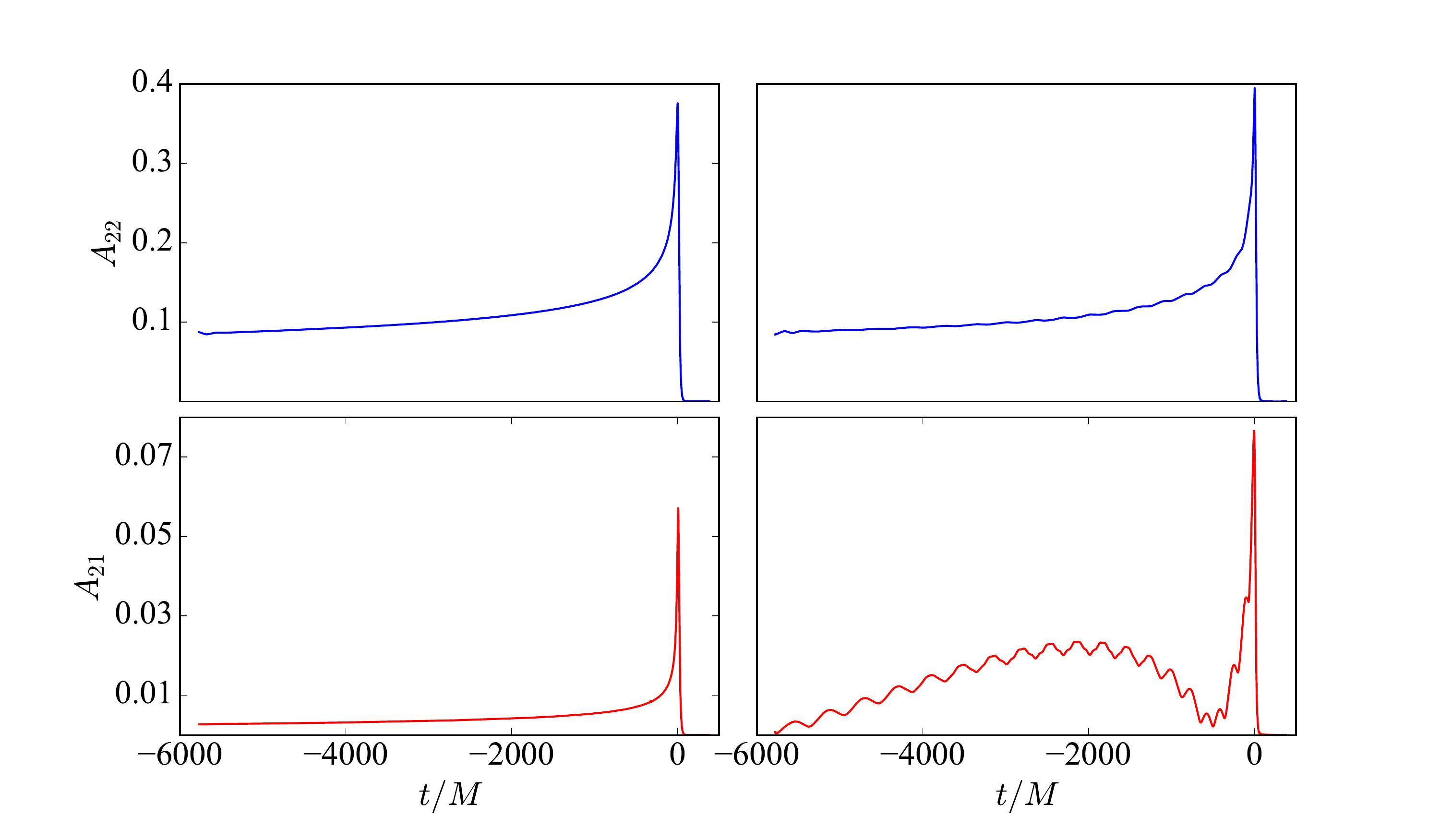}
\caption{
	{\bf Left}: The column shows the amplitudes of the $(2,2)$-mode (top panel) and the $(2,1)$-mode (bottom panel) for the aligned NR case 	\texttt{SXS:BBH:0019}. 
	{\bf Right}: The column shows the same GW modes for the precessing NR configuration \texttt{SXS:BBH:0006}, which shows clear oscillatory features. In both cases, the initial burst of junk radiation has been removed and the modes were aligned such that the peak of the waveform occurs at $t=0$.}
\label{fig:modes}
\end{figure}

We explicitly demonstrate the efficacy of the reduced-order spline method for two distinct BBH configurations. The first once is a case where the (dimensionless) spins $\vec{\chi}_{1,2}$ of the two black holes are (anti-)aligned with the orbital angular momentum $\vec{L}$. The second case is one where the spins 
have some arbitrary initial orientation that causes the orbital plane and spins to precess~\cite{Apostolatos:1994mx, Kidder:1993zz}. 
The GWs produced by aligned-spin and precessing binaries show qualitatively different features, 
which provides a stringent test for the applicability of the reduced-order spline method to a variety 
of binary configurations. While the amplitude and phase of the GWs emitted by aligned-spin 
binaries increase monotonically up to merger, the waveforms of precessing binaries exhibit strong 
amplitude and phase modulations, as illustrated in Fig.~\ref{fig:modes} for the NR cases we consider 
in this analysis. 

The configurations we analyse in detail are two NR simulations from the publicly available SXS waveform 
catalogue~\cite{Mroue:2013PRL}: the aligned-spin configuration \texttt{SXS:BBH:0019}, and the precessing configuration
\texttt{SXS:BBH:0006}. These waveforms were obtained from evolutions with the 
Spectral Einstein Code (SpEC)~\cite{spec, Scheel2014, Szilagyi:2014fna} and are publicly available at~\cite{SXSCatalog}. 
The initial parameters of the numerical simulations are listed in Table~\ref{tab:param}. 
Since for binary black hole simulations the total mass $M=m_1 + m_2$ of the system serves 
as an overall scale, the NR waveform data are output at time samples $\{t_i\}_{i=1}^P$ in units 
of $M$ and can easily be rescaled to any desired total mass in physical units\footnote{Throughout this section we use geometric units and set $G=c=1$.}. 

The NR waveforms are provided as complex gravitational-wave modes $h_{\ell m}(t)$, which are 
obtained by decomposing the time-dependent gravitational radiation field $h(t)$ in a basis of spin-weighted spherical 
harmonics ${}^{s}Y_{\ell m}$ with spin weight $s=-2$,
\begin{equation}
\label{eq:modes}
h_{\ell m}(t) = \int_{S^2} h(t)  {}^{s}Y^*_{\ell m}(\theta, \phi) d\Omega,
\end{equation}
where $(\theta, \phi)$ denote the spherical coordinates on the unit sphere and ${}^*$ denotes complex conjugation.

While all higher harmonics up to and including $(\ell = 8)$ are 
provided by SpEC, we restrict our analysis to the $(\ell, m)$-modes equal to $(2,2)$ and $(2,1)$. The former is the dominant mode and the latter tends to show much stronger modulations in the presence of precession, providing an even more stringent test for the efficacy of the ROS algorithm.

\begin{table}
\centering
\begin{tabular}{c  c c  c  c}
  \hline\hline
  SXS ID & Type & $m_1/m_2$ & $\vec{\chi}_1$ & $\vec{\chi}_2$ \\
  \hline
   \texttt{SXS:BBH:0019} & aligned & 1.5 & $(0.0, 0.0, -0.4995)$ & $(0.0, 0.0, 0.4995)$ \\
   \texttt{SXS:BBH:0006} & precessing & 1.345 & $(0.25, 0.12, -0.16)$ & $(0.1, 0.05, -0.1)$\\
   \hline\hline
\end{tabular}
  \caption{The initial parameters of the binary black hole NR simulations used in this paper. $m_1/m_2$ denotes the mass ratio of the binary, where $m_1$ is the mass assigned to the heavier black hole. $\vec{\chi}_1$ and $\vec{\chi}_2$ denote the initial dimensionless spin vectors $\vec{\chi}_i = \vec{S}_i/m_i^2$ for $i=1,2$ in the Cartesian coordinates of the NR simulation.}
  \label{tab:param}
\end{table}

NR waveform data often contain an initial burst of spurious \emph{junk radiation} as a result of imperfect initial data~\cite{Bowen:1980yu, Damour:2000we, Pfeiffer:2005jf, Garcia:2012dc}, which is removed before the data are subjected to our analysis. 
In addition, certain NR data from the SXS collaboration have been found to contain spurious drifts of the binary's (Newtonian) centre-of-mass due to a large $z$-component of the initial linear momentum. 
This causes the gravitational energy to shift between different $h_{\ell m}$-modes and to induce additional oscillatory behaviour in the waveform modes~\cite{Boyle:2015nqa}, which may hinder the hybridisation process (see below).
To remove these unphysical artefacts, the NR data used in our analysis are subjected to 
Bondi-Metzner-Sachs transformations following the description in~\cite{Boyle:2015nqa} and 
implemented in the publicly available code package {\sc scri}~\cite{scri}. 
More recent SXS simulations use improved initial data~\cite{Ossokine:2015yla} which do no longer show this effect. 

In addition to the relatively short NR waveforms, we created data sets for artificially extended \emph{hybrid waveforms} by attaching a post-Newtonian (PN) inspiral waveform to the NR merger waveforms~\cite{Ajith:2007qp, Santamaria:2010yb, Hannam:2010ky, Ajith:2012az, Schmidt:2012rh, Boyle:2013nka}. 
Since hybrid waveforms are also of interest in GW data analysis, we also demonstrate the efficacy of the the reduced-order spline method for those. 
For the cases analysed, we use a specific PN approximant known as \emph{SpinTaylorT1} as implemented in the publicly available code package {\sc GWFrames}~\cite{gwframes}. 
We choose this specific approximant as the T1-approximants have previously been found to yield better agreement with SXS waveforms over other Taylor approximants~\cite{Scheel:2014ina, Szilagyi:2015rwa}. 
The hybrids have been constructed to reach a starting frequency of 28Hz for a minimal total mass of the binary of $4 M_\odot$.

\subsection{Spline compression}
\label{sec:compression}
For each NR and NR-PN hybrid waveform, we separately analyse the time-domain amplitudes $A_{\ell m}(t)$ and phases $\phi_{\ell m}(t)$ of the $(2,2)$- and $(2,1)$-modes where
\begin{equation}
\label{ }
A_{\ell m}(t) = |h_{\ell m}(t)|, \quad \phi_{\ell m}(t) = \mathrm{arg}(h_{\ell m}(t)).
\end{equation}
Without loss of generality, we apply a time shift to the time-series data such that the peak of the amplitude of the full waveform defined as $h_\mathrm{peak} := \max{(\sum_{\ell, m} h_{\ell m}(t))}$ occurs at $t=0$. 
We choose a fifth degree polynomial ($p=5$) for the spline interpolation order, set the tolerance of the greedy algorithm to be $\epsilon=10^{-6}$, and use the default seed choice as discussed in Sec.~\ref{sec:method}. 
These default choices define our \emph{fiducial splines} as in Sec.~\ref{sec:method}. 
For each of the one-dimensional waveform mode data sets we construct fiducial ROS interpolants following Alg.~\ref{alg:greedy}. 
We determine the reduction of the data size via the reduced-order spline method for the fiducial splines as well as for splines with a tolerance of $\epsilon = 10^{-4}$. In addition, we also investigate how the compression changes when a \emph{relative error measure} is used instead of an \emph{absolute} one to build the ROS. 
To compute the relative greedy error, we replace step 6 in Alg.~\ref{alg:greedy} with 
\begin{equation}
\label{eq:rel}
\vec{\Delta}_{n} = \{ | y_i - S_{n}^{(p)}(x_i) | \}_{i=1}^P \quad
\longrightarrow \quad  \vec{\Delta}_{n} = \bigg\{ \bigg| \frac{ y_i - S_{n}^{(p)}(x_i)}{ \max |\vec{y}| } \bigg| \bigg\}_{i=1}^P  .
\end{equation}

Since we split the waveform into amplitude and phase compared to the original complex $h_{\ell m}$-modes, the total compression per waveform mode $C(h_{\ell m})$ is given in terms of the individual compression factors $C(A_{\ell m}) = {\rm size}(A_{\ell m}) / {\rm size}(A_{\ell m}^{\rm ROS})$ and $C(\phi_{\ell m}) = {\rm size} (\phi_{\ell m}) / {\rm size} (\phi_{\ell m}^{\rm ROS})$ by
\begin{align}
	C(h_{\ell m}) \equiv \frac{ {\rm size}(A_{\ell m}) + {\rm size} (\phi_{\ell m}) }{ {\rm size}(A_{\ell m}^{\rm ROS}) + {\rm size} (\phi_{\ell m}^{\rm ROS}) } = \frac{ 2 C (A_{\ell m}) C(\phi_{\ell m}) }{ C(A_{\ell m}) + C(\phi_{\ell m})}  ,
\label{eq:compression1}
\end{align}
where we have used that the original amplitude and phase data have the same sizes\footnote{
Here, ``size'' can refer to either the number of elements in the data set or the physical memory size on a computer.
}, i.e., ${\rm size}(A_{\ell m}) = {\rm size}(\phi_{\ell m})$.

The compression factors we find are listed in Table~\ref{tab:compressionNR} for the NR waveform modes and in Table~\ref{tab:compressionH} for the hybrids, respectively. 
When the absolute error measure is used with $\epsilon = 10^{-6}$, we observe a greater size reduction for the amplitudes than the phases, in both the NR and the hybrid cases. 
We notice that the lowest data compression factors are found for the precessing NR-PN hybrid case, in particular for those modes which show strong amplitude and phase modulations. This is expected since a more complex data morphology requires the greedy algorithm to retain more points of the original data set to achieve the requested spline accuracy. Nevertheless, even for the $(2,1)$-phase of the precessing hybrid waveform, we achieve a compression of a factor of 2. In addition, we find that the ROS method reduces the physical storage size from $1.1$MB for the $(2,1)$-mode to less than $400$kB. In comparison, for the highest mode compression, which is achieved for the $(2,2)$-mode of the aligned-spin hybrid waveform, we find a size reduction from $1.1$MB to $26$kB. 

The ROS method applied to waveforms which show strong modulations generally results in lower compression factors and,
in general, we find that the overall compression is highly dependent upon the desired spline accuracy $\epsilon$. Changing the accuracy from its default value to $10^{-4}$ significantly increases the compression. For our considered hybrid cases, the aligned-spin $(2,2)$-mode can be reduced to only 8.9kB in digital storage. 

While fixing the tolerance to $\epsilon = 10^{-6}$, even greater mode compression can be achieved when the \emph{relative error} measure in evoked. For all modes considered here, we find that the compression is increased by at least a factor of two in comparison to the absolute error measure, although we find the opposite for lower amplitude higher harmonics  where numerical noise tends to have a larger effect as we will discuss in detail in Sec.~\ref{sec:discussion}.
For the remainder of this section we will be using the absolute error measure to achieve the requested spline accuracy unless stated otherwise.
We note that the numbers quoted should only be interpreted as representative for similar systems. Overall we find that the size of one-dimensional data sets with very little morphology (e.g., aligned-spin waveform modes) can be reduced dramatically via reduced-order spline interpolation. This is particularly true for very long hybrid waveforms, where very few data points need to be retained to build an accurate spline interpolant.

\begin{figure}
\begin{center}
\includegraphics[width=0.49\columnwidth]{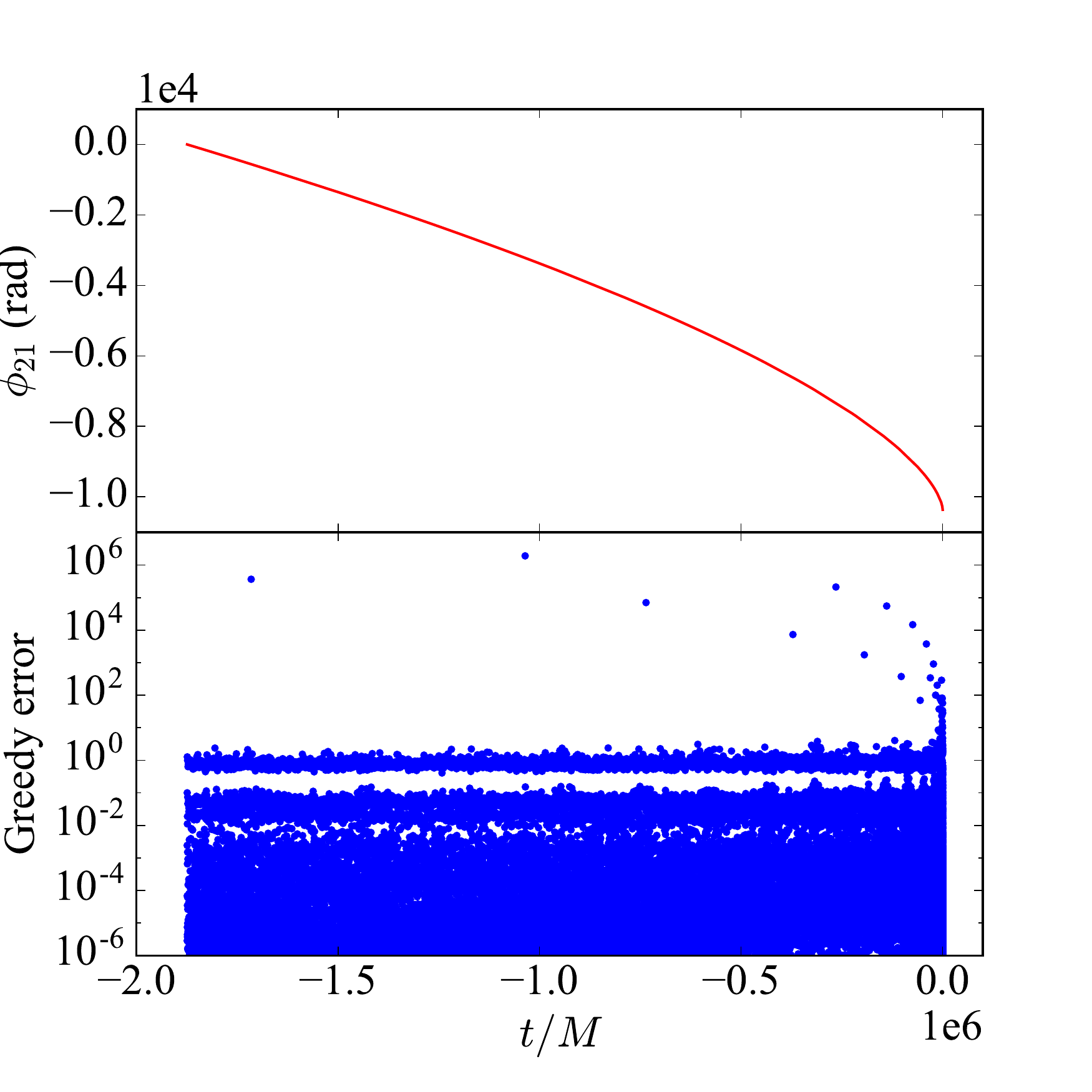}
\includegraphics[width=0.49\columnwidth]{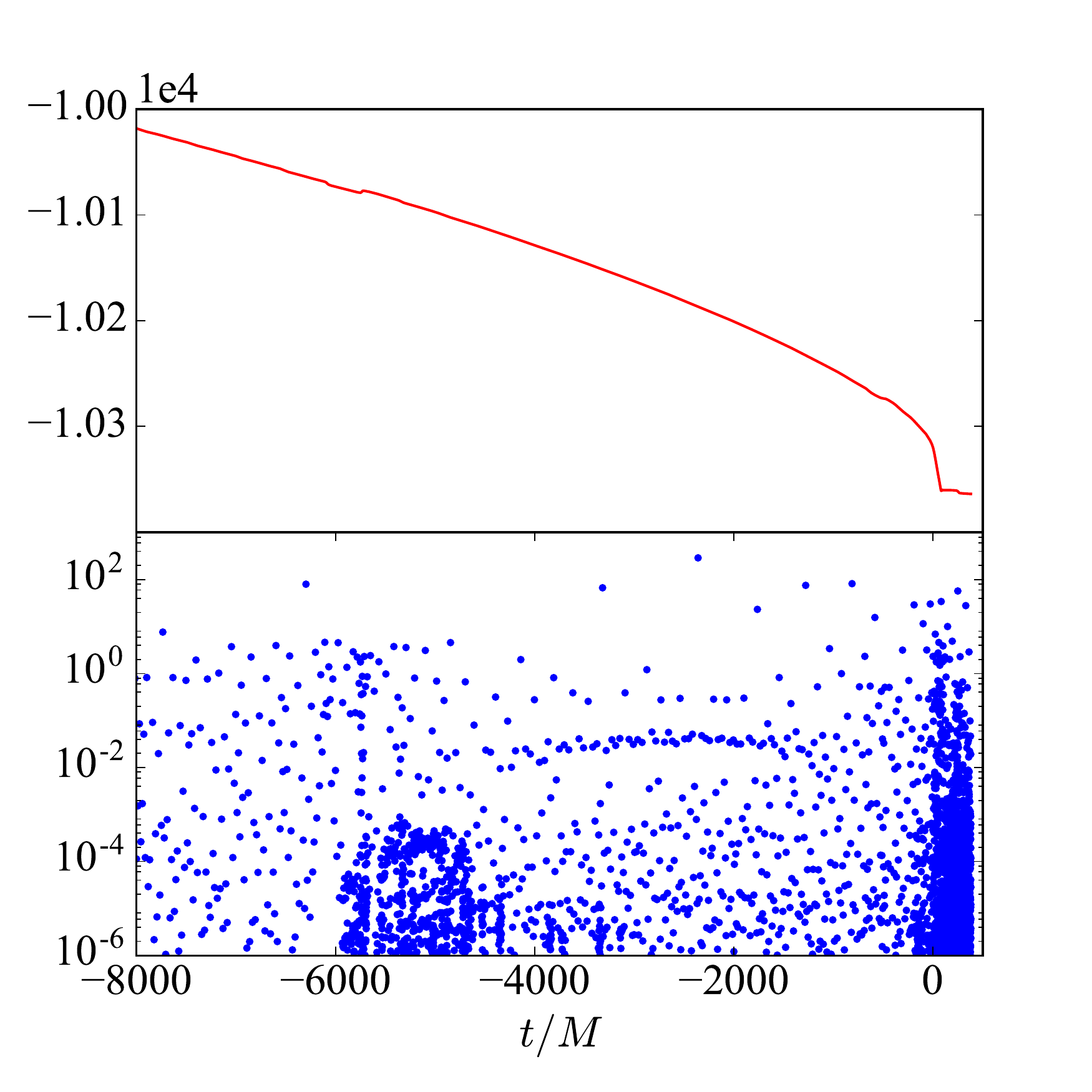}
\caption{
	{\bf Left}: The top panel shows the phase of the $(2,1)$-mode for the precessing NR-PN hybrid 
	waveform. The bottom left panel shows the greedy error versus the corresponding time subsamples selected by the greedy 
	algorithm. 
	{\bf Right}: The panels show the same as the left but from $-8000M$ to  
	ringdown. Significant clustering is seen in the hybridisation window ($t \sim-5000M$) and in the 
	merger-ringdown regimes ($t \gtrsim 0$).}
\label{fig:knots6}
\end{center}
\end{figure}

\begin{figure}
\begin{center}
\includegraphics[width=0.49\columnwidth]{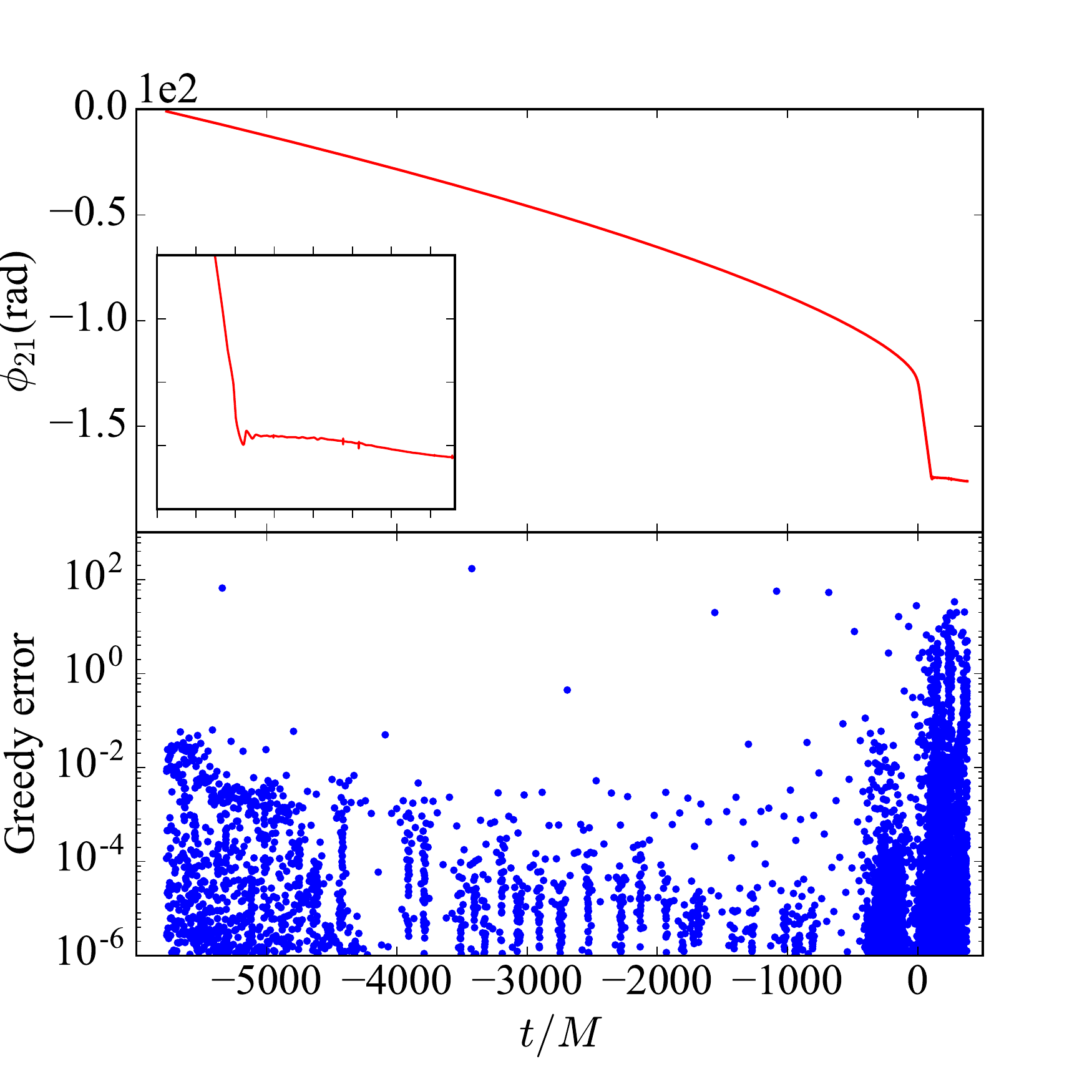}
\includegraphics[width=0.49\columnwidth]{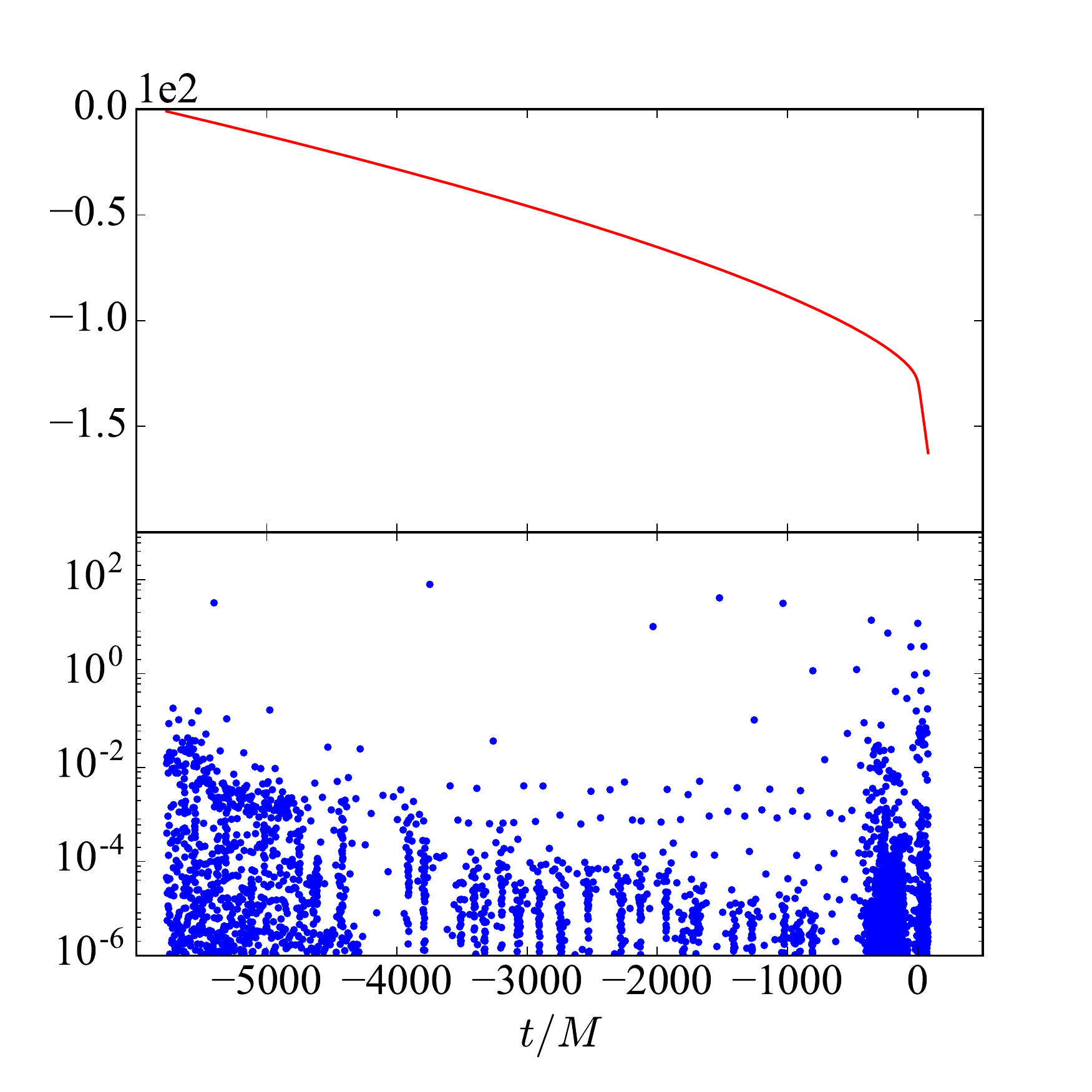}
\caption{
	{\bf Left}: The top panel shows the phase of the $(2,1)$-mode for the nonprecessing NR 
	waveform \texttt{SXS:BBH:0019}. The bottom left panel shows the greedy error versus the corresponding  
	samples selected by the greedy algorithm. The inset in the top panel shows a zoom-in of the ringdown 
	portion, clearly showing the noisy features from numerical errors. 
	{\bf Right}: The panels show the same as the left but 
	when truncating the input data once the numerical noise level is reached after the ringdown. 
	The greedy algorithm is then applied to the truncated data yielding the selected time 
	subsamples in the bottom right panel.}
\label{fig:knots19}
\end{center}
\end{figure}

In addition, looking more closely at the distribution of points selected by the greedy algorithm, we identify various clusters which correspond to morphological features in the waveform data. We visualise the distribution of selected time samples in the fiducial spline for the $(2,1)$-phase of the precessing hybrid waveform in Fig.~\ref{fig:knots6}. Overall, $50\%$ of the original points are selected, but as the right panel shows, we find clusters of a high density. One of the most significant clusters is located in the hybridisation area between $t \sim -6000M$ and $t \sim -4600M$. This is not unexpected as the blending between the NR and the PN waveforms leads to some additional non-smooth structure in the transition region (see also Sec.~\ref{sec:noisy}). The second cluster is found in in the merger-ringdown part of the NR waveform ($t \geq 0$). While we expect an increased density around the peak time due to steep changes in the shape of the phase function, the high accumulation of points at late times in the ringdown are an indication for numerical noise in the waveform data. The greedy algorithm tries to resolve this noise to the prescribed accuracy which can only be achieved by selecting almost all original data points. We illustrate this point in more detail by looking at aligned-spin NR case \texttt{SXS:BBH:0019}. Fig.~\ref{fig:knots19} shows the distribution of selected time samples for the ROS of the $(2,1)$-phase with (left panel) and without the numerical noise in the late ringdown stage (right panel). Removing this part of the waveform data \emph{before} applying the ROS method results in a slightly improved compression factor of $C(h_{21})=9.6$ instead of 7.0.

\begin{table}
  \centering 
  \begin{tabular}{c | c | c | c || c | c | c }
	\hline\hline
	& \multicolumn{3}{c}{0019 NR} & \multicolumn{3}{c}{0006 NR}  \\
	Compression & $\mathrm{abs}^{10^{-6}}$ & $\mathrm{abs}^{10^{-4}}$ & $\mathrm{rel}^{10^{-6}}$ & $\mathrm{abs}^{10^{-6}}$ & $\mathrm{abs}^{10^{-4}}$ & $\mathrm{rel}^{10^{-6}}$\\
	\hline
   $C(A_{22})$ & 29.1 & 156.0 & 18.5 & 30.4 & 131.5 & 24.7 \\
   $C(\phi_{22})$ & 7.3 & 21.8 & 27.5 & 8.1 & 27.2 & 40.0 \\
   $C(h_{22})$ & 11.7 & 38.3 & 22.1 & 12.8 & 45.1 & 30.5 \\
   \hline
   $C(A_{21})$ & 28.3 & 333.1 & 9.6 & 30.1 & 95.1 & 17.3 \\
   $C(\phi_{21})$ & 4.0 & 12.4 & 14.0 & 5.6 & 17.1 & 23.0 \\
   $C(h_{21})$ & 7.0 & 23.9 & 11.4 & 9.5 & 29.0 & 19.7 \\
   \hline \hline
\end{tabular}
  \caption{The compression factors of the fiducial splines for the $(2,2)$ and $(2,1)$ amplitudes and phases of both NR data sets. $C(h_{\ell m})$ indicates the achieved total $h_{\ell m}$-mode compression as given in Eq.~\eqref{eq:compression1}. We list the compression factors for the absolute error measure (abs) and the relative error measure (rel), as well as for different spline tolerances ($\epsilon=10^{-6}$ and $\epsilon=10^{-4}$).} 
  \label{tab:compressionNR}
\end{table}

\begin{table}
  \centering 
  \begin{tabular}{c | c | c | c || c | c | c }
	\hline\hline
	& \multicolumn{3}{c}{0019 Hybrid} & \multicolumn{3}{c}{0006 Hybrid}  \\
	Compression & $\mathrm{abs}^{10^{-6}}$ & $\mathrm{abs}^{10^{-4}}$ & $\mathrm{rel}^{10^{-6}}$ & $\mathrm{abs}^{10^{-6}}$ & $\mathrm{abs}^{10^{-4}}$ & $\mathrm{rel}^{10^{-6}}$\\
	\hline
   $C(A_{22})$ & 104.3 & 455.0 & 66.9 & 10.5 & 26.0 & 9.0 \\
   $C(\phi_{22})$ & 25.6 & 71.3 & 285.1 & 5.3 & 12.6 & 103.2 \\
   $C(h_{22})$ & 41.4 & 123.3 & 108.4 & 7.0 & 17.0 & 16.6 \\
   \hline
   $C(A_{21})$ & 110.2 & 874.0 & 35.0 & 5.5 & 11.6 & 3.7 \\
   $C(\phi_{21})$ & 13.9 & 44.2 & 151.3 & 2.0 & 4.1 & 7.9 \\
   $C(h_{21})$ & 24.7 & 84.1 & 56.9 & 2.9 & 6.1 & 5.0 \\
   \hline \hline
\end{tabular}
  \caption{The compression factors of the fiducial splines for the $(2,2)$- and $(2,1)$-amplitudes and phases of both PN-NR hybrid data sets. $C(h_{\ell m})$ indicates the achieved total $h_{\ell m}$-mode compression as given in Eq.~\eqref{eq:compression1}. We list the compression factors for the absolute error measure (abs) and the relative error measure (rel), as well as for different spline tolerances ($\epsilon=10^{-6}$ and $\epsilon=10^{-4}$).} 
  \label{tab:compressionH}
\end{table}

\subsection{Greedy seed choice and polynomial degree}
As discussed in Sec.~\ref{sec:seed1} for our test function, the choice of initial seeds for the greedy algorithm affects the size of the resulting spline interpolant. To estimate the average obtainable compression rate, we repeat the greedy point selection algorithm a thousand times on each one-dimensional phase and amplitude data set using random choices of seeds in comparison to the default seed choice. For all analysed data sets we find that 
while the default seed choice does not necessarily produce the highest possible compression for a given tolerance $\epsilon$ and polynomial degree $p$, the fiducial spline sizes lie well within the $3\sigma$-interval of the average size.

\begin{figure}
\begin{center}
\includegraphics[width=\columnwidth]{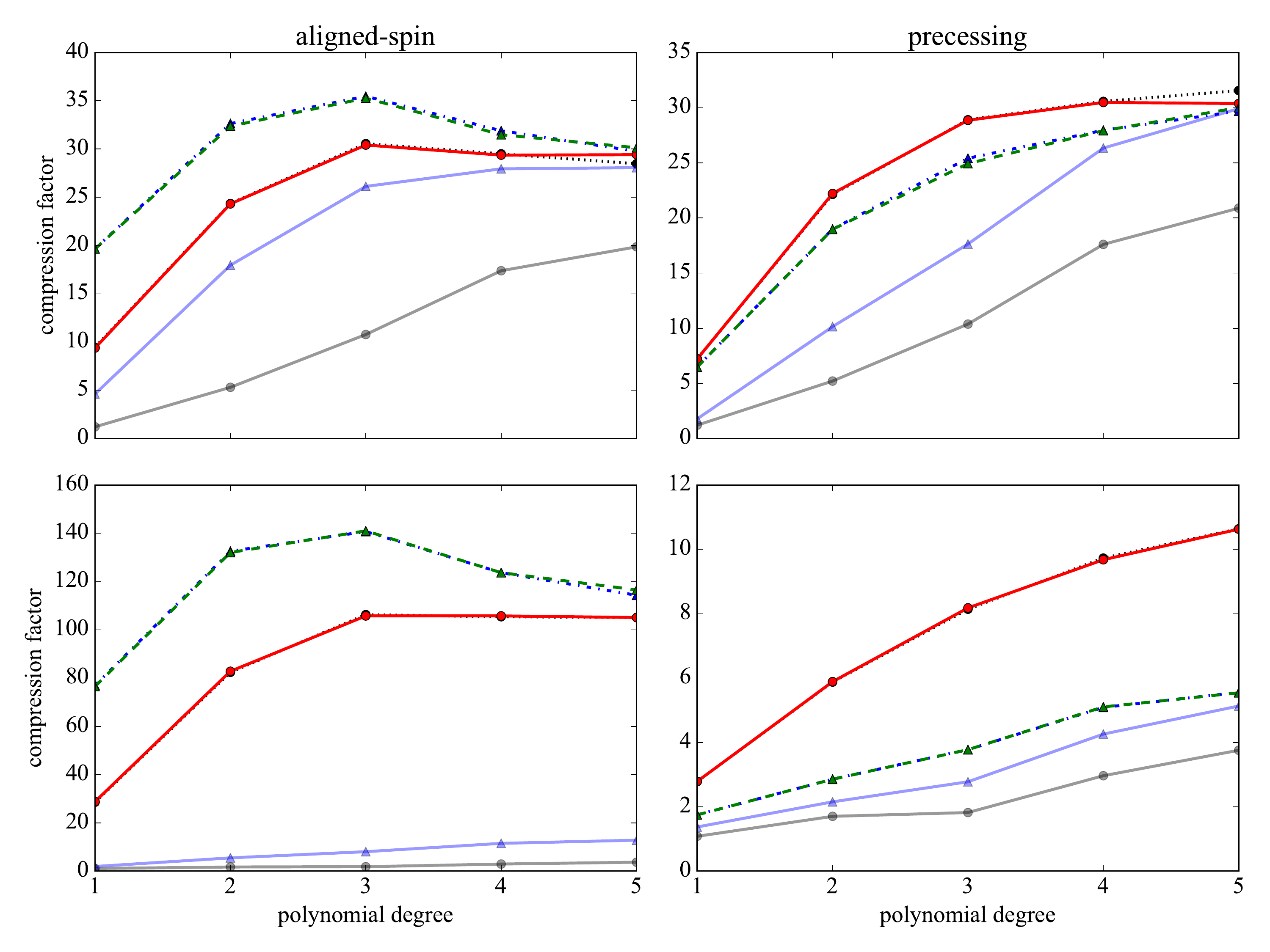}
\caption{
	The four panels show the \emph{maximal compression factors} for the amplitudes, phases and the real parts of the $(2,2)$- and $(2,1)$-modes for the  aligned-spin case (left) and the precessing case (right) as function of the polynomial degree of the spline. The curves for the imaginary parts are barely distinguishable from the results of the real parts.
	{\bf Top}: The top panel shows  the results for the unhybridised NR cases: The red solid curve and the black dotted curve represent the amplitude and phase compression of the $(2,2)$-mode. The blue dot-dashed and the green dashed curves show the same quantities for the $(2,1)$-mode. The two solid lower curves show the maximal compression for the real parts of the $(2,2)$-mode (circles, grey) and the $(2,1)$-mode (triangles, blue).
	{\bf Bottom}: The bottom panels show the same for the hybridised versions. The maximal compression factors are obtained by making 100 different choices to seed the greedy algorithm, only the highest compression factors per polynomial degree are shown. For the default choice of a fifth degree polynomial, the compression factors for amplitudes, phases as well as each mode, are listed in Table~\ref{tab:compressionNR} and Table~\ref{tab:compressionH}. When using the real and imaginary decomposition, we find the  following total compression factors for the fifth degree polynomial: 20.3 and 27.9 (aligned), 21.3 and 29.1 (precessing), 3.7 and 12.8 (aligned hybrid), 3.8 and 5.1 (precessing hybrid). We obtain almost identical results for the imaginary parts.}
\label{fig:smallest}
\end{center}
\end{figure}

In addition to the choice of initial seeds, for a fixed default tolerance $\epsilon=10^{-6}$, we find that the polynomial degree $p$ of the spline interpolant impacts the maximal compression rate that can be achieved. We illustrate this effect in Fig.~\ref{fig:smallest} for both unhybridised NR cases as well as for their corresponding hybrids. For each one-dimensional data set and polynomial degree $p = 1,2,3,4,5$, we choose one-hundred different realisations of random initial seeds. In comparison to the compression factors obtained for our fiducial ROS as listed in Table~\ref{tab:compressionNR} and Table~\ref{tab:compressionH}, we see immediately that the smallest reduced data size is not necessarily found for the default seed choice. However, for all analysed data sets we find that our default choice for the polynomial degree, $p=5$, yields either the highest or only slightly lower compression factors than cubic or quartic polynomials, also shown in Fig.~\ref{fig:smallest}. We note that for the aligned-spin case $p=3$ results in the highest compression, which we attribute to the noisy features in the late ringdown, where quintic polynomials may be prone to overfitting the data (also see discussion in Sec.~\ref{sec:compression}).

Instead of decomposing the GW modes into amplitude and phase, one could also consider the real and imaginary part of each mode and investigate whether higher compression factors can be obtained in comparison to the amplitude and phase decomposition. To do so, we repeat the previous analyses using the real and imaginary parts of the $h_{\ell m}$-modes. The best results obtained for the real parts of the waveforms for one-hundred realisations of initial greedy seeds as a function of polynomial degree are also shown in Fig.~\ref{fig:smallest}: We find comparable compression factors for the pure NR cases (top panel), but the ROS method is significantly less efficient for the hybrid waveforms when the real and imaginary parts are used (bottom panel). We attribute this to the oscillatory structure of the real and imaginary parts, which requires more points to achieve the targeted spline tolerance compared to the amplitude and phase decomposition. We obtain almost identical results for the imaginary parts.

In addition, care needs to be taken with \emph{memory modes}, i.e. $(\ell, 0)$. 
Commonly used methods to extract gravitational waveforms in numerical simulations may retain gauge-related artefacts that render memory modes either unphysical or inaccurate~\cite{Taylor:2013zia}. 
For waveforms from aligned-spin binaries, the memory modes are purely real but often contain small imaginary pieces due to the presence of numerical errors in a NR simulation. 
As such, a decomposition into amplitude and phase is inappropriate for these modes and a naive application of the ROS method to this case produces very little compression of the phase, as anticipated from the results in Sec.~\ref{sec:noisy}.
However, there is no obstacle to applying the ROS method to the well-defined real and imaginary components of the memory modes, which leads to significantly improved compression.
For waveforms from precessing binaries, the memory modes are generally complex. 
As such, a decomposition into either amplitude and phase or real and imaginary pieces is perfectly appropriate for these modes but due to their possibly small magnitude, the data themselves may have significant numerical noise content and, therefore the real-imaginary decomposition may be more favourable.

\subsection{Monte Carlo $K$-fold cross-validation}
For applications of reduced-order spline interpolation to NR waveforms in data analysis, we are particularly interested in assessing the quality of the interpolant at points that  
were not contained in the original (non-uniform) time series $h_{\ell m}(t)$. 

In GW data analyses, waveform data need to be re-sampled uniformly at a certain sampling rate $f_s$, which determines the times at which the spline interpolant will be evaluated. By construction, the ROS only reproduces the original waveforms to within the specified accuracy $\epsilon$. Once a sampling rate $f_s$ and a total mass $M$ are chosen, the spline is evaluated at the new time array. To assess the uncertainty of the ROS in predicting the values of the waveforms at those new times, we perform a Monte Carlo $K$-fold cross-validation study. As described in Sec.~\ref{sec:cvTest}, this allows one to estimate the mean error of the spline in predicting new values. 

\begin{figure}
\includegraphics[width=0.5\columnwidth]{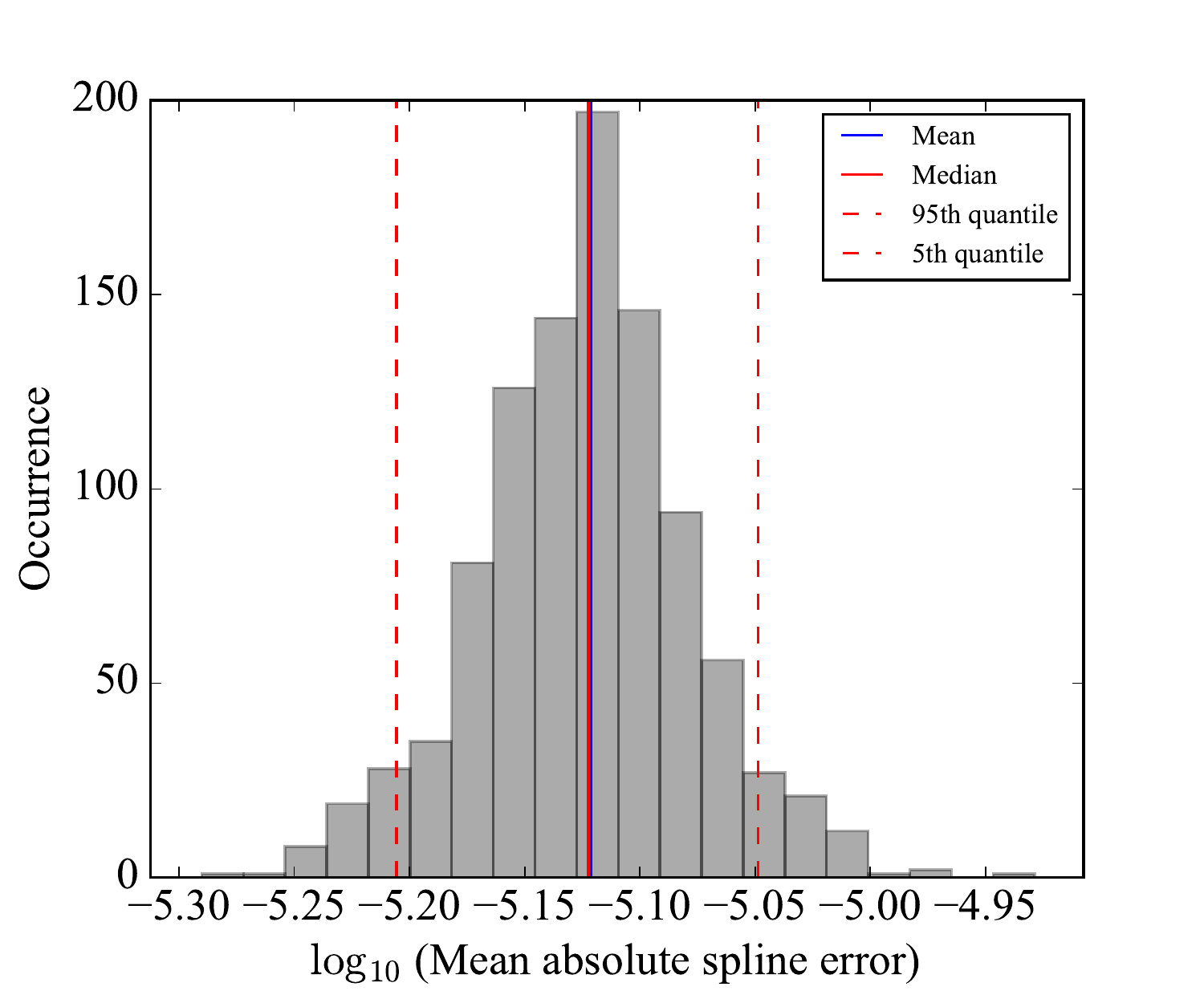}
\includegraphics[width=0.5\columnwidth]{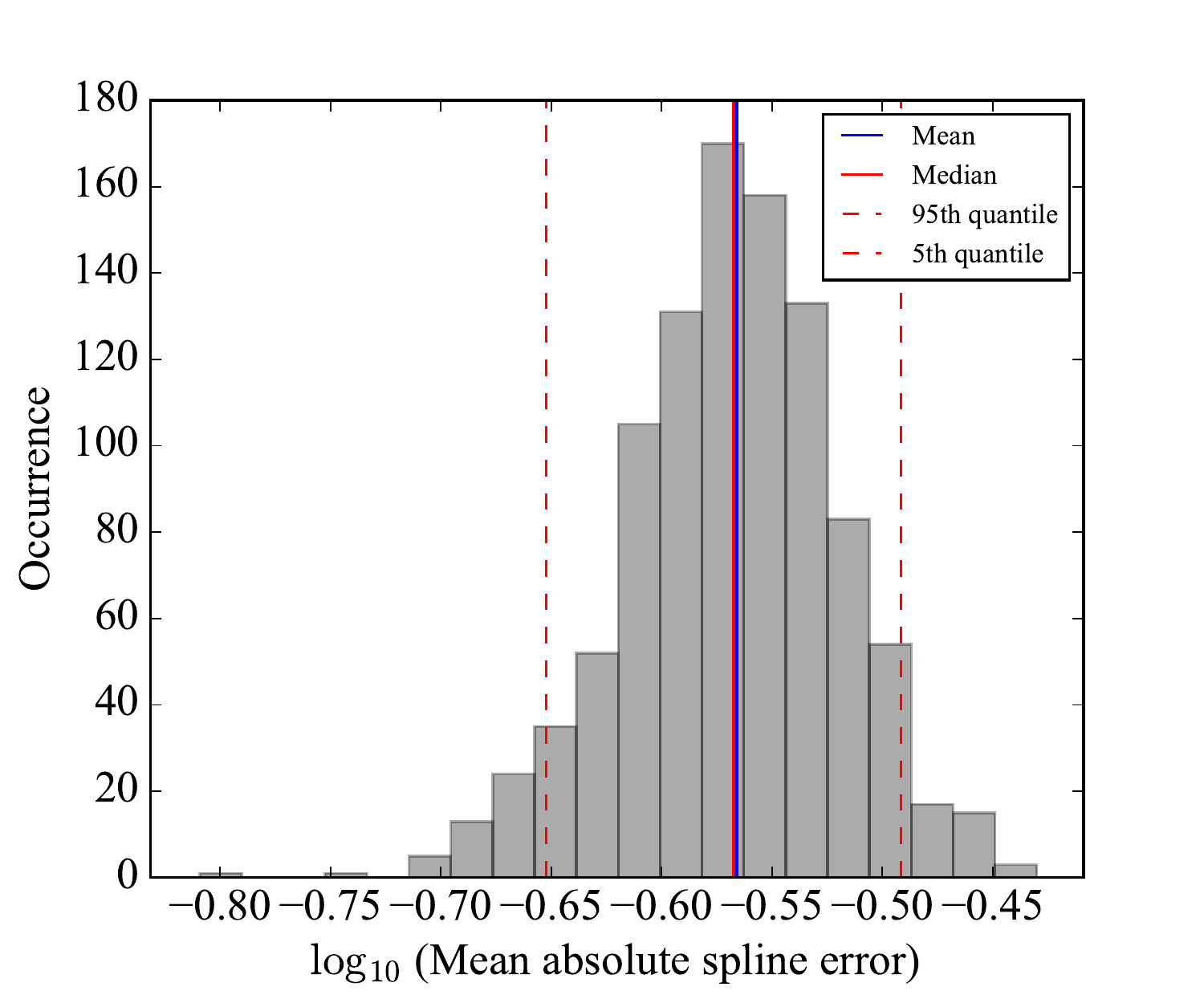}
\caption{
	{\bf Left}: The panel shows the mean absolute spline error obtained from the $K$-fold cross validation study for the amplitude of the $(2,1)$-mode of 	the NR case \texttt{SXS:BBH:0019}. 
	{\bf Right}: This panel shows the results for the phase of the same mode.}
\label{fig:Kfold19}
\end{figure}

For each one-dimensional data set we perform one thousand $K$-fold cross-validations as discussed in Sec.~\ref{sec:cvTest}. 
For the NR cases we find that on average the mean absolute spline error in the amplitude is of the order of $10^{-5}$. For the phases, however, we find that the errors can be as large as $\sim 10^{-1}$ as illustrated in Fig.~\ref{fig:Kfold19}. As noted in the previous section, we find that the density of the points selected by the greedy algorithm is particularly high at the very end of the NR data (see left panel of Fig.~\ref{fig:knots19}). At this stage, the ringdown has subsided and the data only contain numerical noise. By construction, however, the ROS interpolant tries to resolve this highly-oscillatory noise, which can only be achieved by including almost all points in this part of the waveform (see Sec.~\ref{sec:noisy}). Due to the noisy character, predicting these points when excluded from the training set is extremely difficult, hence the relatively poor performance in the cross-validation. We have repeated this analysis by excluding the noisy data after the ringdown from the data set. We find a mean absolute spline error of $\leq 10^{-5}$ for the $(2,2)$-phase and $\leq 10^{-3}$ for the phase of the $(2,1)$-mode. The mean errors in the amplitudes only decrease marginally. 
For the hybrid cases, we obtain comparable values to the NR cases. If, however, the amplitude-phase mode decomposition is replaced by the real and imaginary parts of each waveform mode, we consistently find average spline errors of $\leq 10^{-5}$ in the cross-validation studies suggesting that noise contained in the NR part of the hybrid data is pronounced in the amplitude-phase decomposition and is indeed the cause of the increased mean spline error.

\subsection{ROS vs. linearly interpolated data}
To quantify whether the ROS method impacts the quality of numerical/hybrid GW waveforms for data analysis applications
in comparison to previously used methods \footnote{Linear interpolation of NR/hybrid data was used previously in the NINJA and NINJA-2 projects~\cite{Aylott:2009ya,Ajith:2012az,Aasi:2014tra}}, we compute the \emph{overlap} between the individual waveform modes obtained from the linear waveform interpolants and from the ROS interpolants. We show that no additional loss of waveform accuracy is introduced by using the ROS method instead of linear interpolation. 

The overlap is commonly defined as the normalised noise-weighted inner product between two waveforms where no optimisation over time or phase shifts are performed. It is a frequently used measure in GW data analysis to indicate how similar two waveforms are and therefore serves as an ideal measure to identify whether the ROS interpolant produces a less accurate representation of the NR/hybrid data than linear interpolation. 
The calculation is conveniently performed in the Fourier domain, and
since we compute the overlap between two complex GW modes $h_{\ell m}$, we will be using the following definition of the overlap:
\begin{equation}
\label{eq:match}
\mathcal{O} =  \frac{\langle h_1(t) | h_2(t) \rangle}{ \sqrt{ \langle h_1(t) | h_1(t) \rangle \langle h_2(t) | h_2(t) \rangle} },
\end{equation}
where
\begin{equation}
\label{ }
\langle h_1(t) | h_2(t) \rangle := \mathrm{Re}\bigg[ \bigg( \int_{f_{\rm min}}^{f_{\rm max}} + \int _{-f_{\rm max}} ^{-f_{\rm min}} \bigg) \frac{\tilde{h}_1^*(f) \tilde{h}_2(f)}{S_n( |f| )}  df\bigg].
\end{equation}
Here, $h_i(t)$ denotes the $i$-th complex time-domain waveform mode, $\tilde{h}_i(f)$ its Fourier transform, $S_n(|f|)$ is the one-sided power spectrum and ${}^*$ denotes complex conjugation. The integration is performed over \emph{all} frequencies since the waveforms from precessing binaries generally have spectral content in both, positive and negative frequencies. 
The motivation for this particular definition and how it relates to the complex norm of the difference between two modes is given in~\ref{sec:appA}.

\begin{figure}
\begin{center}
\includegraphics[width=0.49\textwidth]{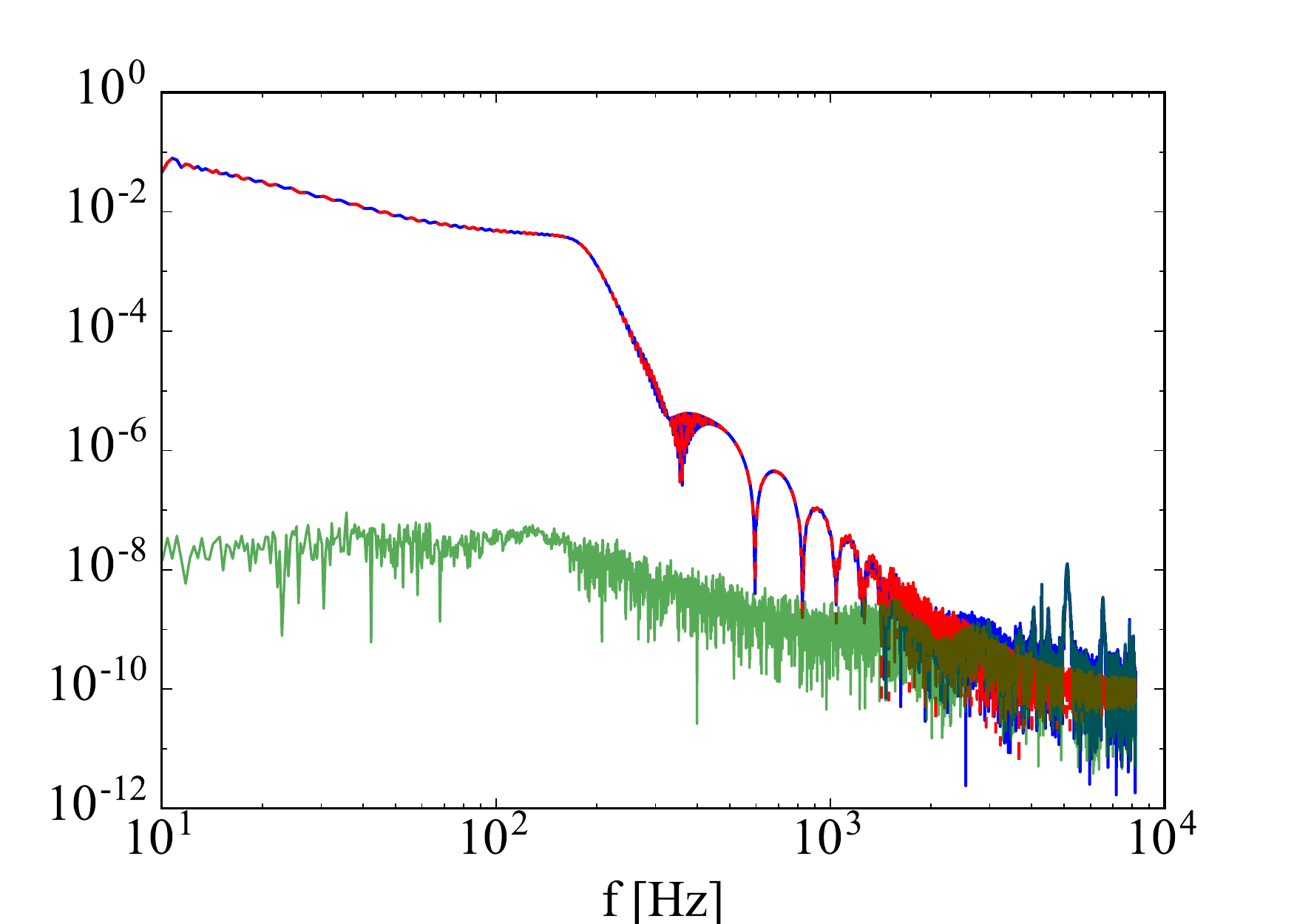}
\includegraphics[width=0.49\textwidth]{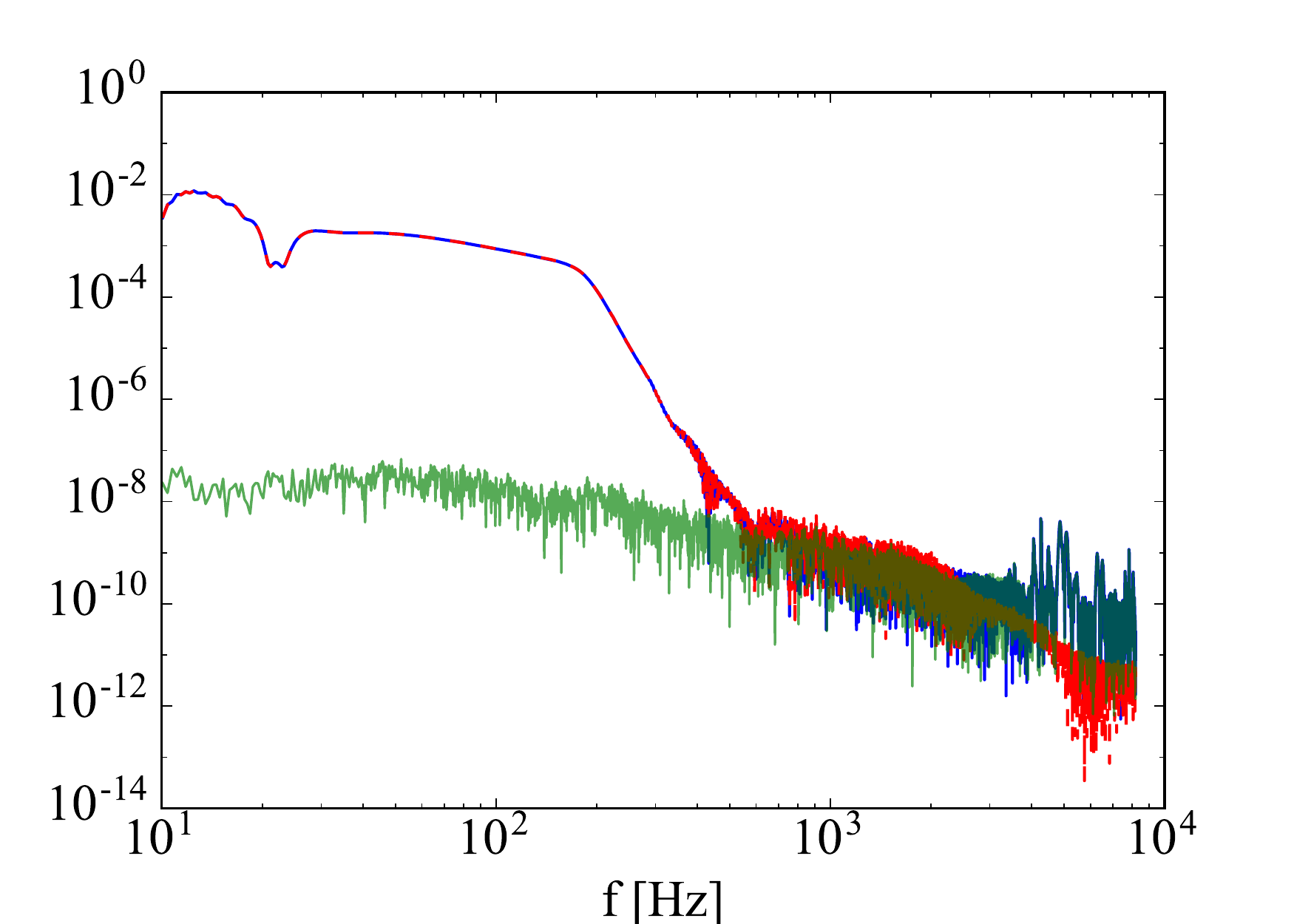}
\caption{{\bf Left}: The panel shows the Fourier domain amplitude of the $(2,2)$-mode of the precessing NR waveform. We only show the positive frequencies starting at 10Hz. The modes are sampled at $f_s=16,384$Hz for a total mass of $95M_\odot$. The blue graph depicts the waveform obtained from the linear interplant, the red (dashed) graph the ROS. {\bf Right}: The same as in the left panel but now for the $(2,1)$-mode. In both panels the green curve shows the difference between the curves.  As anticipated, in both cases the differences between the curves are negligibly small. We also note a slight increase in difference at very high frequencies, which we attribute to the noisy features in the NR data after the ringdown has subsided as discussed in Sec.~\ref{sec:compression}.}
\label{fig:FDwave}
\end{center}
\end{figure}

\begin{figure}
\begin{center}
\includegraphics[width=\textwidth]{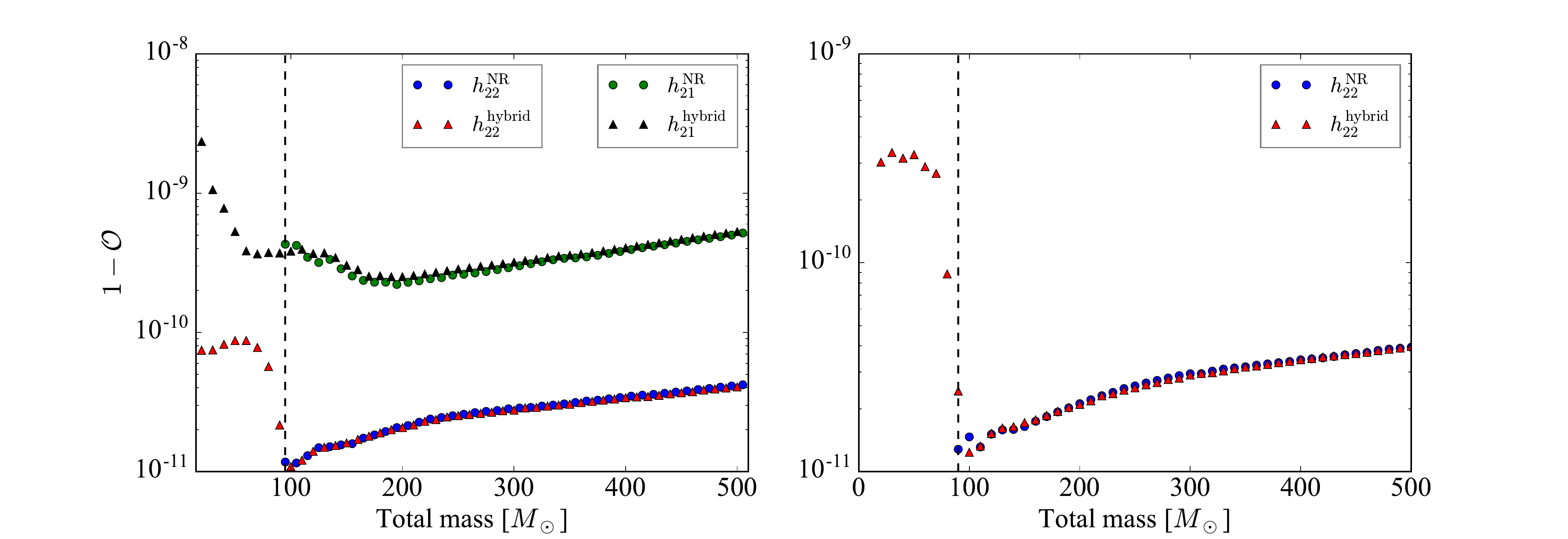}
\caption{{\bf Left}: The panel shows the disagreement, $1-\mathcal{O}$, between the linear interpolant and the corresponding ROS for the aligned-spin NR and hybrid waveforms for the $(2,2)$- and the $(2,1)$-modes. {\bf Right}: This panel shows the same quantity for the $(2,2)$-mode of precessing NR and hybrid case. The vertical dashed line indicates the minimal total mass for the NR waveforms.}
\label{fig:overlap}
\end{center}
\end{figure}

To quantify the agreement between the linearly interpolated NR/hybrid data and the ROS, we compute the overlap between the individual $h_{\ell m}$-modes under consideration and therefore are not concerned with a more general overlap computation, which is required when multiple modes are considered at the same time~\cite{Capano:2013raa, Schmidt:2014iyl}. To avoid any edge effects from the Fourier transform due to the band-limitation of the waveforms, we apply a Planck taper~\cite{McKechan:2010kp} to the time domain waveforms. 
Fig.~\ref{fig:FDwave} shows the (positive-frequency) Fourier domain amplitudes of the $(2,2)$- and $(2,1)$-mode of the precessing NR waveform obtained by linear interpolation (blue) and via the ROS method (red dashed).
Fig.~\ref{fig:overlap} shows the overlap for the $h_{22}$- and $h_{21}$-modes between the linearly interpolated and their corresponding ROS for both the NR and the hybrid data. We compute the overlaps over a range of total masses $M \in [20, 500]\,M_\odot$ with a waveform starting frequency of $f_\mathrm{min}=10$Hz and a sampling rate of $f_s = 16,384$Hz. We use the Advanced LIGO design sensitivity power spectral density (zero-detuned high power)~\cite{noise}. 

For both cases, the aligned-spin (left panel) and the precessing one (right panel), we find that the differences are of the order $\leq 10^{-9}$ for the $(2,2)$-mode and $\leq 10^{-8}$ for the $(2,1)$-mode, with slightly better agreement in the aligned-spin case. We also note that the disagreement between the linear and ROS interpolant is larger for low-mass systems. For all cases and modes we find that the difference in accuracy between the ROS and linear interpolant is negligible and does not impact the waveform accuracy for data analysis. 

We have repeated this comparison for a selection of total masses when the relative error measure is used instead of the absolute one to build the ROS of the phase. In that case we find that $1 - \mathcal{O} \leq 10^{-6}$, directly reflecting the spline tolerance $\epsilon$.   

\subsection{A new gravitational-wave data format}
The analyses presented in the previous sections have shown that the reduced-order spline method indeed provides an accurate representation of a one-dimensional NR or hybrid data set like a gravitational-wave mode $h_{\ell m}$. We have shown that no additional losses in accuracy are incurred in comparison to traditionally used linear interpolation, making the ROS method equally suitable for data analysis applications. 
One tremendous advantage of the presented ROS method is the significant reduction of the data load for any large data set, but in particular for multimodal gravitational waveforms. 

As a result of our analyses we conclude that the reduced-order spline interpolation provides an excellent method to efficiently compress and accurately represent NR and NR-PN hybrid waveform data. We therefore suggest to use this new \emph{compressed waveform format} to store waveform data and
to enable fast and accurate incorporation of NR/hybrid waveforms into GW data analysis pipelines. 

This compressed waveform format is an integral part of the new LIGO NR injection infrastructure~\cite{nrinj}, which is fully implemented in LAL~\cite{lal}. It has already been used extensively to complement various studies related to the analysis of LIGO's first GW detection, the binary black hole merger GW150914, using Numerical Relativity waveforms~\cite{GW150914, TheLIGOScientific:2016src, TheLIGOScientific:2016wfe, TheLIGOScientific:2016uux, TheLIGOScientific:2016abc, Abbott:2016izl, Lovelace:2016uwp}. The ROS representation of NR and hybrid waveforms can be generated via the publicly available Python module \textsc{romSpline}~\cite{romspline}.

\section{Discussion}
\label{sec:discussion}
We have presented, characterised and implemented a greedy algorithm for determining a relevant subset of
a one-dimensional data set that is sufficient for representing the original data by a spline interpolant to some requested 
point-wise accuracy. The reduced data set is recovered by the ROS to numerical round-off errors by 
construction. The remaining data are reproduced up to the requested tolerance 
threshold used in the greedy algorithm to construct the interpolant.
The resulting size of the reduced data set depends on the structure of the data itself with more 
compression occurring for data with fewer features. 
Data containing noise with small amplitudes can be compressed with a ROS up to the noise level. We caution, however, that the interpolation of noisy features is mathematically problematic since interpolation fundamentally relies on continuity and smoothness, neither of which typically manifests in noisy data.

The resulting reduced-order spline can then be used to resample the data where necessary.
We have also discussed and performed cross-validation studies to assess the interpolation errors 
and the robustness of the greedy algorithm to changes in input parameters. Cross-validation
is also useful for revealing portions of the data that have noisy or non-smooth features that may be due to actual 
noise or undesirable structure/artefacts in the data. 

To demonstrate the properties and efficiency of the ROS method for application to gravitational waves, in Sec.~\ref{sec:GW} we applied the greedy algorithm to NR and NR-PN waveforms, focusing only on the $(2,2)$- and $(2,1)$-modes for illustration purposes. We have performed  various analyses to demonstrate the robustness of the method when applied to GW modes, and find that the maximal achievable compression depends on the choice of decomposition, the polynomial degree $p$ of the spline interpolant and the spline tolerance $\epsilon$. With this method, we are also able to assign a median error in predicting new values to the interpolant, which we have demonstrated via Monte Carlo $K$-fold cross-validation. With this, for the NR waveforms we have identified undesirable features during the last stages of ringdown, where the simulations are not 
resolving the small amplitudes very well. Such data is likely not necessary for most 
applications and one can simply remove that data before implementing the greedy algorithm.

Further, we have shown that for data analysis purposes, the linear interpolants of the full original data
sets and the corresponding ROS interpolants are indistinguishable. While the accuracy of the interpolation method is 
comparable, the ROS method is far superior when it comes to the required data storage due to the greedy compression. 

However, the full data sets do not only contain the $(2,2)$- and $(2,1)$-modes, but in fact contain \emph{all} radiative modes ($\ell \ge 2$) up to a maximum of $\ell =8$. 
When applying the ROS method to \emph{all} 77 available modes for each NR and NR-PN case discussed in Sec.~\ref{sec:GW}, we find that, 
depending on the particular $(\ell, m)$-mode, the greedy algorithm may produce a smaller reduced data set when applied to the amplitude and phase instead of the real and imaginary parts, or vice versa. We find a strong correlation between the numerical mode resolution and the decomposition. The smaller in magnitude the mode, the more numerical noise is prevalent and the decomposition into real and imaginary part is often preferred. 
We also find that, as $\ell$ increases, the mode amplitudes generally decrease in magnitude so that there tend to be more compact reduced datasets for a fixed absolute tolerance. Conversely, the reduced datasets tend to be less compact when using a relative tolerance as the greedy algorithm becomes more sensitive to the increasingly prevalent numerical noise.
We have applied the greedy algorithm to all 77 available modes of the NR and NR-PN data discussed here, storing to file the reduced data for the smaller of the amplitude-phase and real-imaginary representations of the mode. 
In this way, we find that the original NR data files get significantly compressed when the most optimal representation per $(\ell, m)$-mode is chosen, which is shown in Table~\ref{tab:compression11}.

\begin{table}
  \centering 
  \begin{tabular}{c | c  c  | c  c | c  c | c  c }
  \hline\hline
    & \multicolumn{8}{c }{File size (MB)} \\
    $\epsilon$  & \multicolumn{2}{c}{0019 NR}  & \multicolumn{2}{c}{0006 NR}  & \multicolumn{2}{c}{0019 hybrid} &  \multicolumn{2}{c}{0006 hybrid}   \\
    \hline
     N/A & \multicolumn{2}{c |}{26.5} & \multicolumn{2}{c |}{26.2} & \multicolumn{2}{c |}{83.5} & \multicolumn{2}{c}{85.1} \\
	 & abs & rel & abs & rel & abs & rel & abs & rel \\
\hline
   $10^{-3}$ & 1.2 & 2.8 & 1.2 & 2.2 & 1.2 & 2.3 & 1.6 & 3.2 \\
   $10^{-4}$ & 1.2 & 4.8 & 1.3 & 3.5 & 1.4 & 3.8 & 2.4 & 6.9 \\
   $10^{-5}$ & 1.4 & 7.5 & 1.5 & 5.9 & 2.0 & 6.3 & 4.1 & 14.4 \\
   $10^{-6}$ & 2.0 & 10.9 & 2.0 & 9.2 & 3.1 & 9.5 & 7.1 & 24.0 \\
   \hline \hline
\end{tabular}
  \caption{File sizes (in MB) for the NR and hybridised NR-PN data sets compressed using the ROS greedy algorithm for various threshold tolerances $\epsilon$ in both absolute (abs) and relative (rel) measures.  
  The file sizes of the original data are shown in the first line. All 77 modes of the original data 
  are compressed. The smaller of the amplitude-phase and real-imaginary reduced data sets are saved to disk. Hence, the
  smallest file size for all modes is obtained by a mix of amplitude-phase and real-imaginary representations.} 
  \label{tab:compression11}
\end{table}

The greedy algorithm for building a reduced-order spline is implemented in the public Python 
code {\sc romSpline} available for download at~\cite{romspline}. 
Included are tutorials showing how to use the code in the context of the example presented in Sec.~\ref{sec:method}.
We use the {\it UnivariateSpline} 
class in the {\it scipy.interpolate} module to generate our spline interpolants because of the code's 
speed and because one has direct access to the first $p$ derivatives of the spline predictions, which 
may be useful for other applications. However, in such applications an even more accurate ROS may be generated by first 
computing the desired derivatives numerically (e.g., using finite differencing) and then applying the 
greedy algorithm.

In order to use the NR injection infrastructure in the publicly available LIGO Algorithms Library, the NR/hybrid
data have to be provided in the compressed waveform format. The details on how to build these data sets accordingly
can be found in~\cite{nrinj}.

We have shown that the reduced-order spline algorithm Alg.~\ref{alg:greedy} presented in this paper provides an efficient method to compress 
large relatively smooth, one-dimensional data sets while obtaining an accurate interpolant of the uncompressed data set, 
allowing for fast evaluation and resampling of the data set as desired while significantly reducing the required storage. 

\section*{Acknowledgements}
We are grateful to Scott Field and Kent Blackburn for very helpful comments and suggestions on a previous draft of the paper.
The authors also thank Ian Harry, Daniel Hemberger, Geoffrey Lovelace, Harald Pfeiffer, Geraint Pratten, Mark Scheel, and Alan Weinstein for useful discussions and comments. 
C.R.G.~and P.S.~were supported by the Sherman Fairchild Foundation and NSF grant PHY-1404569 to the California Institute of Technology. C.R.G. also thanks the Brinson Foundation for partial support. P.S.~was also supported by NSF grant PHY-1151197 to the California Institute of Technology and acknowledges support from the LIGO Laboratory and the National Science Foundation. LIGO was constructed by the California Institute of Technology and the Massachusetts Institute of Technology with funding from the National Science Foundation and operates under cooperative agreement PHY-0757058. \\

\appendix

\section{Overlap between complex GW modes}
\label{sec:appA}
Given two complex gravitational waveforms, $h_1(t), h_2(t) \in \mathbb{C}$, we define a symmetric inner product by
\begin{equation}
\label{ }
\langle h_1(t) | h_2(t) \rangle := \int_{-\infty}^{\infty} \frac{\tilde{h}_1(f) \tilde{h}_2^*(f)}{S_n(|f|)} df
\end{equation}
The complex norm of the difference between such two waveforms is then related to the overlap in the following way:
\begin{align}
\label{}
||h_1(t) - h_2(t)||^2 &\equiv \langle h_1(t) - h_2(t) | h_1(t) - h_2(t) \rangle   \\
                              &= \int_{-\infty}^{\infty} \frac{| \tilde{h}_1(f) - \tilde{h}_2(f)|^2}{S_n(|f|)} df \\
                              &= \int_{-\infty}^{\infty} \frac{df}{S_n(|f|)}\big\{ \tilde{h}_1\tilde{h}_1^* + \tilde{h}_2 \tilde{h}_2^* -\tilde{h}_1\tilde{h}_2^* - \tilde{h}_1^*\tilde{h}_2\big\} \\
                              &= 2 - \int_{-\infty}^{\infty} \frac{df}{S_n(|f|)}\{ \tilde{h}_1\tilde{h}_2^* + \tilde{h}_1^*\tilde{h}_2\} \\
                              &= 2 - 2 \mathrm{Re} \langle h_1(t) | h_2(t) \rangle  \\
                              &= 2(1 - \mathcal{O})
\end{align}
\\
\section*{References}
\bibliographystyle{iopart-num}
\bibliography{rom}

\end{document}